\documentclass[aps,10pt,prd,twocolumn,preprintnumbers,amsmath,amssymb,nofootinbib,groupedaddress,a4paper,showpacs]{revtex4-1}
\usepackage{amsmath,amssymb}
\usepackage{xcolor}
\usepackage[colorlinks=true, linkcolor=blue, urlcolor=blue, citecolor=blue, anchorcolor=blue]{hyperref}     
\usepackage{graphicx}
\usepackage{algorithmicx}
\usepackage{algpseudocode}
\usepackage{algorithm}
\algrenewcommand\algorithmicindent{0.4em}

\DeclareMathOperator\Tr{Tr}
\DeclareMathOperator\qr{qr}
\newcommand{\erfc}{{\rm erfc}}
\newcommand{\ex}[1]{\left\langle #1 \right\rangle}
\newcommand{\var}[1]{{\rm var}\left[#1\right]}
\newcommand{\npf}{n_{\mathrm{pf}}}
\newcommand{\nf}{{N_{\rm f}}}
\newcommand{\ns}{{N_{\rm shifts}}}
\newcommand{\norm}[1]{{\left|{#1}\right|}}
\newcommand{\Sf}{{S^{\rm eff}_{\mathrm{f}}}}
\newcommand{\Sg}{{S_{\mathrm{g}}}}
\def\cpp{{{C\nolinebreak[4]\hspace{-.05em}\raisebox{.4ex}{\tiny\bf ++}}}}

\graphicspath {{figures/}}

\begin{document}

\preprint{CERN-TH-2018-161}

\title{Rational Hybrid Monte Carlo with Block Solvers and Multiple Pseudofermions}
\author{Philippe de Forcrand}
\email{forcrand@phys.ethz.ch}
\author{Liam Keegan}
\email{keeganl@phys.ethz.ch}
\affiliation{Institut f\"ur Theoretische Physik, ETH Z\"urich, CH-8093 Z\"urich, Switzerland}
\date{\today}

\begin{abstract}
The dominant cost of most lattice QCD simulations is the inversion of
the Dirac operator required to calculate the force term in the RHMC
update. One way to improve this situation is to use multiple pseudofermions,
which reduces the size and variance of this force and hence allows a larger
integration step size to be used. This means fewer force term calculations are required, but at the cost of having to invert the
Dirac operator for each pseudofermion field. This bottleneck can be addressed: recently there has been renewed interest in the use of block Krylov solvers, which can solve multiple right hand side vectors with significantly fewer iterations than are required if each vector is
solved using a separate Krylov solver. We combine these two ideas, achieving a significant speed-up of RHMC lattice QCD simulations.
\end{abstract}


\maketitle

\section{Introduction}
\label{sec:intro}
The main difficulty in lattice simulations of QCD is calculating the determinant 
of the Dirac operator, a very large and badly conditioned matrix. In the Rational Hybrid Monte Carlo~\cite{Kennedy:1998cu,Clark:2003na,Clark:2006wq} (RHMC) approach, this determinant is stochastically estimated by inverting the Dirac operator acting on a bosonic field of ``pseudofermions'' using an iterative Krylov solver. The RHMC evolution requires the numerical integration of the pseudofermion force term, and when this term is large or has a large variance a small integrator step size must be used, resulting in many costly pseudofermion force calculations.

Many different approaches have been proposed to reduce the computing
cost of RHMC. They range from preconditioning the solver (e.g. even-odd~\cite{DeGrand:1990dk,LIPPERT19991357}, domain decomposition~\cite{Luscher:2003qa,Frommer:2012zm}, deflation~\cite{Luscher:2007se}, multigrid~\cite{Frommer:2013fsa,Brower:2018ymy}) to preconditioning the action (ILU \cite{deForcrand:1996ck},
UV-filtering \cite{deForcrand:1998sv}) to tuning the integrator 
(\cite{Takaishi:2005tz,Clark:2008gh,Kennedy:2012gk}).

In particular, a popular strategy which reduces the RHMC fermionic force term is the ``Hasenbusch trick'' or ``mass splitting'', and its generalisations~\cite{Hasenbusch:2001ne,Hasenbusch:2002ai}. One replaces the Dirac matrix $M$ by $(M H^{-1}) H$, where $H$ is associated with a heavy fermion, and represents each of the two determinants by a pseudofermion integral. The value of the heavy mass can be tuned to minimise the computer cost per accepted Hybrid Monte Carlo (HMC) trajectory. This tuning becomes more challenging in the case of multiple mass splittings; an empirical rule consists of adjusting the magnitude of the pseudofermion forces to be the same for each factor.

A simple way to obtain a similar effect is to replace $M$ with $\left[M^{\tfrac{1}{\npf}}\right]^{\npf}$~\cite{Clark:2006fx}, and represent each of the $\npf$ determinants by a pseudofermion integral. The resulting force magnitude is automatically the same for all factors, and only one parameter, the number $\npf$ of pseudofermions, needs to be adjusted. The cost is that the Dirac operator must be inverted on $\npf$ pseudofermion vectors for each force term calculation.

Recently there has been renewed interest~\cite{Sakurai:2009rb,Tadano:2009gg,Nakamura:2011my,Birk:2012tn,Birk2014,Birk:thesis,Clark:2017ekr} in the use of block Krylov solvers~\cite{OLEARY1980293}, which invert the same matrix on multiple vectors simultaneously, and thanks to the enlarged Krylov basis from which solutions are constructed, can converge with significantly fewer iterations than are required to solve each vector separately.

Here we combine these two ideas to speed up the RHMC algorithm.

\section{Multiple Pseudofermions}
\label{sec:npf}
The partition function we want to sample, for $\nf$ degenerate--mass quarks, is given by
\begin{equation}
 \mathcal{Z} = \int dU e^{-\Sg} \det \left[M^{\dagger}M\right]^{\nf/2} = \int dU e^{-\Sg -\Sf},
\end{equation}
where $\Sg$ is the gauge action and $M$ the Dirac operator, and both are functions of the gauge field $U$. To sample this using HMC requires the calculation of the fermionic force term,
\begin{equation}
\label{eq:force_exact}
 F^{a}_{x\mu} = 
 -\frac{\partial\Sf}{\partial U^a_{x\mu}}
= \Tr \left[\left(M^{\dagger}M\right)^{-\tfrac{\nf}{2}}\frac{\partial \left(M^{\dagger}M\right)^{\tfrac{\nf}{2}}}{\partial U^a_{x\mu}}\right],
\end{equation}
where $a$ is the color index, $x$ the site index, and $\mu$ the direction index. This would require the entire Dirac operator to be diagonalised. To avoid doing this, the determinant can be written as an integral over bosonic pseudofermion fields $\phi$ which gives (up to an overall constant) the equivalent partition function,
\begin{equation}
 \mathcal{Z} = \int dU d\phi d\phi^{\dagger} e^{-\Sg -\phi^{\dagger} [M^{\dagger}M]^{-\nf/2} \phi},
\end{equation}
where pseudofermions with the desired distribution can be generated by first sampling $\eta$ from a normal distribution, then constructing $\phi = \left[M^{\dagger}M\right]^{\nf/4}\eta$. The fractional powers of $M^{\dagger}M$ acting on a vector can in all cases be approximated to any desired accuracy by use of a suitable rational approximation of the form
\begin{equation}
 \label{eq:RA}
 [M^{\dagger}M]^{r} x \simeq \alpha_0 \, x + \sum_{j=1}^{\ns} \alpha_j (M^{\dagger}M + \beta_j)^{-1} x,
\end{equation}
where the coefficients $\alpha_j,\beta_j > 0$ and the number of shifts $\ns$ depend on the exponent $r$, the spectral range of the Dirac operator, and the desired accuracy of the approximation.

This approach can be extended to multiple pseudofermions; using the trivial identity
  \begin{equation}
    \det \left[M^{\dagger}M\right] = \det \left[\left(M^{\dagger}M\right)^{\tfrac{1}{\npf}}\right]^{\npf},
  \end{equation}
the partition function can instead be written as
\begin{equation}
 \mathcal{Z} = \int dU \prod_{i=1}^{\npf}\left( d\phi_i d\phi_i^{\dagger} \right) e^{-\Sg -\sum_{i=1}^{\npf} \phi_i^{\dagger} [M^{\dagger}M]^{-\frac{\nf}{2\npf}} \phi_i},
\end{equation}
where $\eta_i$ are again sampled from a normal distribution, and $\phi_i = \left[M^{\dagger}M\right]^{\frac{\nf}{4\npf}}\eta_i$.

The resulting pseudofermion force term for $\npf$ pseudofermions is given by
\begin{equation}
  \label{eq:force}
F^a_{x\mu}(\phi_i, U, \npf) = \sum_{i=1}^{\npf} \phi_i^{\dagger} \frac{\partial \left[M^{\dagger}M\right]^{-\frac{\nf}{2\npf}}}{\partial U^a_{x\mu}} \phi_i.
\end{equation}
  
For a given gauge field $U$, writing the $\phi_i$ fields in terms of the gaussian $\eta_i$ fields, then integrating over them in Eq.~(\ref{eq:force}) we recover the correct expectation value of the force term, Eq.~(\ref{eq:force_exact}), which is independent of $\npf$,
  \begin{align}
  \label{eq:force_npf}
    \overline{ F^a_{x\mu}(U,\npf) } & \equiv \int \prod_{i=1}^{\npf}\left( p(\eta_i)d\eta_i \right) F^a_{x\mu}(\left[M^{\dagger}M\right]^{\frac{\nf}{4\npf}}\eta_i,U,\npf) \\ \nonumber
    &= \Tr \left[\left(M^{\dagger}M\right)^{-\tfrac{\nf}{2}}\frac{\partial \left(M^{\dagger}M\right)^{\tfrac{\nf}{2}}}{\partial U^a_{x\mu}}\right],
  \end{align}
with a variance that is suppressed by $\npf$,
  \begin{equation}
  \label{eq:force_var}
    \left[\overline{F^a_{x\mu}(U,\npf)^2}\right] - \left[\overline{F^a_{x\mu}(U,\npf)}\right]^2 = \frac{c_1}{\npf} + \mathcal{O}(\npf^{-2}),
  \end{equation}
where $c_1$ does not depend on $\npf$.
In simulations we can easily measure the norm $F^2$ of this pseudofermion force,
  \begin{equation}
  \label{eq:force_norm}
    F^2(\npf) = \left\langle\sum_{a x \mu}\tfrac{1}{2}{\left[F^a_{x\mu}(\phi_i, U, \npf)\right]^2}\right\rangle,
  \end{equation}
where $\langle\dots\rangle$ represents an average over the gauge fields. Moreover, for the particular choice of the 2nd order Omelyan~\cite{OMELYAN2003272,Takaishi:2005tz} integrator with $\lambda=1/6$, the variance of this norm is related to the variance of the energy violation $\Delta H$~\cite{Bussone:2018mzi} over a trajectory of length $\tau$ with integrator step size $\delta \tau = \tau / n_{\mathrm{steps}}$\footnote{Note that $\tau$ may need to be rescaled if the choice of normalisation of the kinetic term in the HMC differs from that of Ref.~\cite{Bussone:2018mzi}.},
\begin{equation}
  \label{eq:force_dH}
  \var{\Delta H} = 8 \left(\frac{\delta\tau}{12}\right)^4 \var{F^2(\npf)} + \mathcal{O}(\delta\tau^6).
  \end{equation}
This relation is valid up to higher order corrections in the step size, and assumes that the trajectory length is long enough that the correlation between initial and final force terms can be neglected. Here we also assume that a multi--scale integrator~\cite{Sexton:1992nu} is used such that the gauge force term's contribution to the integrator error is negligible.
This variance in the trajectory energy violation can in turn be related to the acceptance $P_{\mathrm{acc}}$ using the Creutz acceptance formula~\cite{Creutz:1988wv,Gupta:1990ka} 
\begin{equation}
  \label{eq:creutz}
  P_{\mathrm{acc}}(\Delta H) 
  = \erfc(\sqrt{\var{\Delta H}/8}),
  \end{equation}
which is valid for high acceptances. Combining the two and expanding in $\delta\tau$ gives the simple prediction for the acceptance,
  \begin{equation}
  \label{eq:p_acc_force}
    P_{\mathrm{acc}} = 1 - \tfrac{1}{72\sqrt{\pi}} \delta\tau^2 \sqrt{\var{F^2(\npf)}} + \mathcal{O}(\delta\tau^4),
  \end{equation}
and assuming that the total trajectory cost is dominated by the force term inversions, the relative cost $C(\npf)$ of simulations at different $\npf$ can be estimated as the cost of a force term inversion ($\propto\npf$) multiplied by the number of inversions ($\propto 1/\delta\tau$),
  \begin{equation}
  \label{eq:cost}
    C(\npf) \propto \npf/\delta\tau \propto \npf \left(\var{F^2(\npf)}\right)^{1/4},
  \end{equation}
which we can use to cheaply estimate the relative performance of simulations using different values of $\npf$ simply by measuring the variance of the force term for each $\npf$ on the same set of thermalised configurations.
Another estimate for the cost is given in Ref.~\cite{Clark:2006fx},
\begin{equation}
\label{eq:cost_clark}
C(\npf) \propto \npf^2 \kappa^{\tfrac{1}{\npf}}
\end{equation}
where $\kappa$ is the condition number of the Dirac operator. We will compare these simple estimates with the actual cost of simulations for different $\npf$ in Sec.~\ref{sec:results}. 
For large values of $\npf$ Eq.~(\ref{eq:force_var}) gives the $\npf$--dependence of the force norm as,
\begin{equation}
  \label{eq:force_norm_npf}
    F^2(\npf) = c_0 + c_1 \npf^{-1} + \mathcal{O}(\npf^{-2})
  \end{equation}
and similarly for the variance of this norm one finds,
  \begin{equation}
  \label{eq:force_norm_var_npf}
    \var{F^2(\npf)} = c_2 \npf^{-1} + c_3 \npf^{-2} + \mathcal{O}(\npf^{-3}),
  \end{equation}
where the constants $c_i$ are expectation values of traces involving the Dirac operator that do not depend on $\npf$, and in particular $c_0 = F^2$ is the norm of the exact force term of Eq.~(\ref{eq:force_exact}).

We see that increasing $\npf$ reduces this variance, which according to Eq.~(\ref{eq:p_acc_force}) will allow a larger step size to be used in the integrator, resulting in fewer force term calculations. The lowest shift $\beta_1$ in the rational approximation of Eq.~(\ref{eq:RA}) also increases with $\npf$, which makes the inversion of the Dirac operator converge faster. These gains are offset by the cost of inverting the Dirac operator $\npf$ times, however empirical studies have shown that using intermediate values of $\npf>1$ result in a smaller total simulation cost than $\npf=1$~\cite{Clark:2006fx}.

In the next section we further improve on this idea, taking advantage of the presence of multiple pseudofermions to reduce the cost of these $\npf$ Dirac operator inversions, by combining the pseudofermion vectors at each site on the lattice to form a block matrix (or ``pencil''). This has two benefits: applying the Dirac operator to the block matrix is more computationally efficient than applying it to each vector in turn, and the block structure allows the use of a block multishift--CG inverter which requires fewer Dirac operator calls to converge.

\section{Block Krylov Solvers}
\label{sec:block}
A Krylov solver iteratively solves the system $A x = b$ for the vector $x$ given some vector $b$, where we take $A$ to be a hermitian positive definite matrix. Starting from some initial guess $x^{(0)}$ with residual $r = b - A x^{(0)}$, it constructs a solution $x^{(k)}$ after $k$ iterations from the Krylov basis $\mathcal{K}_k = \left\{r, Ar, A^2r, \dots, A^{k-1}r \right\}$. The conjugate gradient (CG) solver is an example of such a Krylov solver; at each step it finds the solution that minimises the error norm $\norm{e_k}_A \equiv (x^{(k)}-x^*)^{\dagger} A (x^{(k)}-x^*)$, where $x^*$ is the exact solution.
  
Since we want to solve for $\npf$ vectors $b_j$, where $j=1,2,\dots,\npf$, with the \emph{same} Dirac matrix for each vector, we can form a block matrix $B$ whose $j$-th column is $b_j$, and solve the system $A X = B$. The solution is now constructed from the much larger block-Krylov basis $\mathcal{K}_k = \left\{R, A R, A^2 R, \dots, A^{k-1}R \right\}$, where $R = B - AX^{(0)}$, which can potentially converge with significantly fewer iterations. Additionally there can be a performance gain from only having to read the matrix $A$ once per $\npf$ vectors. Extending the CG solver in this way gives the Block CG (BCG) algorithm~\cite{OLEARY1980293}, which minimises $\Tr \left[(X^{(i)}-X^*)^{\dagger} A (X^{(i)}-X^*)\right]$ at each step, and is equivalent to CG for $\npf=1$.

\begin{figure}
\begin{center}
  \includegraphics[width=25em]{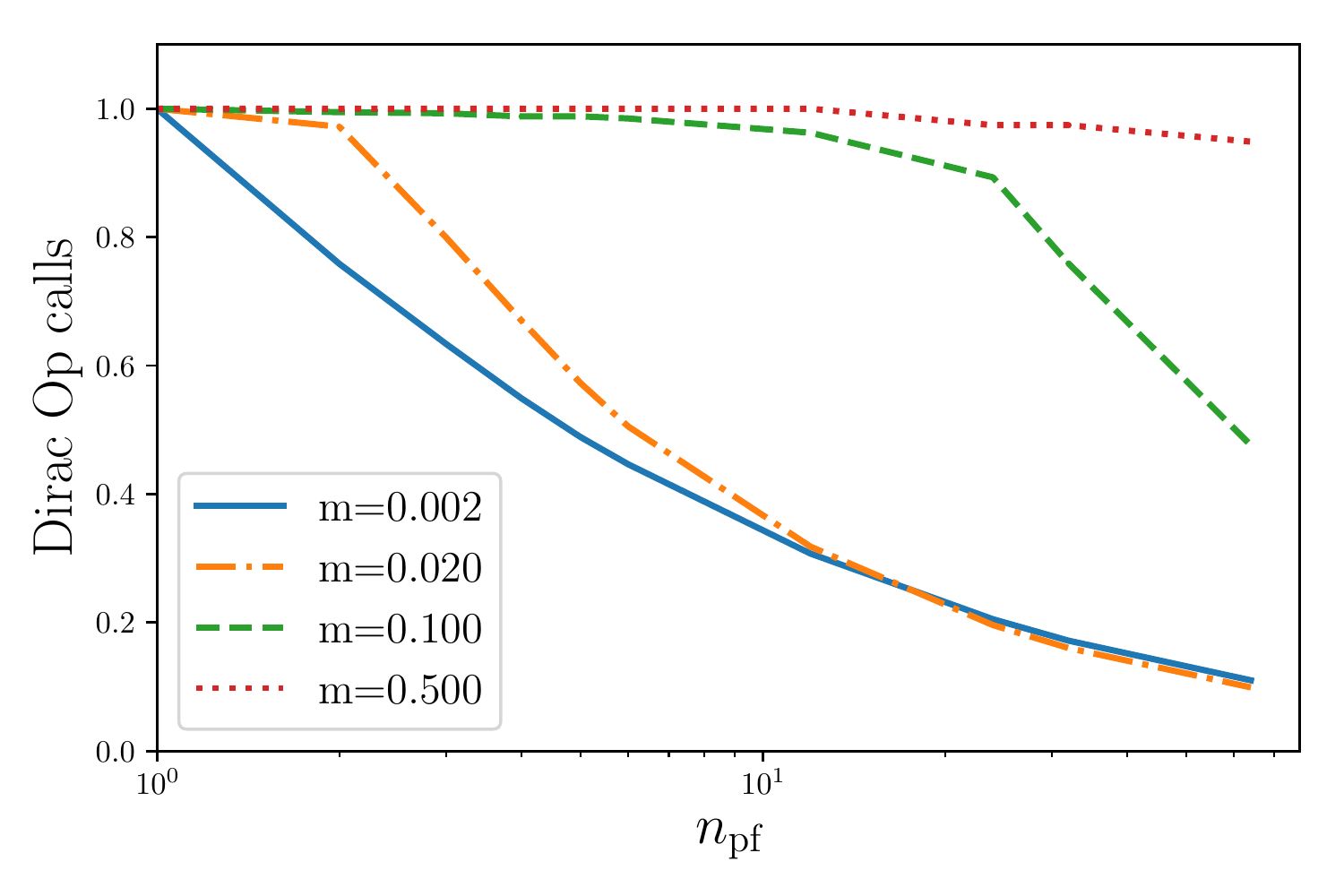}\\\includegraphics[width=25em]{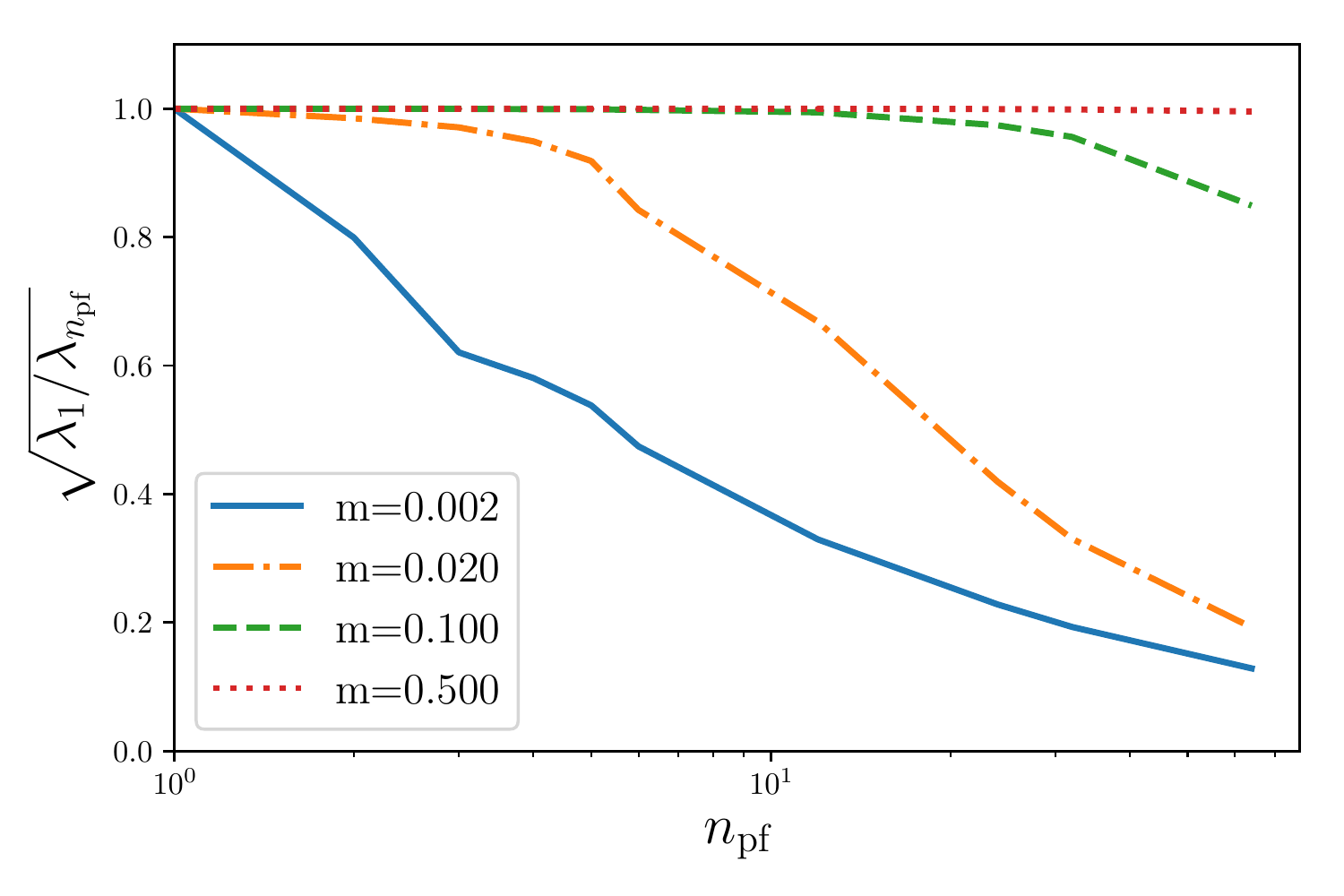}
\caption{\emph{Top}: Number of Dirac operator calls for the block SBCGrQ solver to converge compared to the SCG solver, for a range of fermion masses. As the mass is made lighter, the block solver improvement increases. \emph{Bottom}: The square root of the ratio of the lowest eigenvalue to the $\npf$--th eigenvalue of the Dirac operator. The bound on the convergence rate of Eq.~(\ref{eq:convergence_bound}) is determined by this quantity, and qualitatively it also seems to describe the actual convergence of the block solver quite well.}
\label{fig:convergence}
\end{center}
\end{figure}

There is an upper bound on the relative error of the BCG solution after $k$ steps~\cite{OLEARY1980293},
\begin{equation}
 \frac{\norm{e_k}_A}{\norm{e_0}_A} \leq c_1(\npf) \left(
 \frac{1 - \sqrt{\lambda_{\npf}/\lambda_{\mathrm{max}}}}{1 + \sqrt{\lambda_{\npf}/\lambda_{\mathrm{max}}}}
 \right)^{2k}
\end{equation}
where the eigenvalues of $A$ in ascending order are given by $\left\{\lambda_1, \lambda_2, \dots, \lambda_{\npf}, \dots, \lambda_{\mathrm{max}} \right\}$, and $c_1(\npf)$, where $c_1(1)=4$, is a function that we will approximate as constant here. Expanding in powers of $\sqrt{\lambda_{\npf}/\lambda_{\mathrm{max}}}$ this can be written as
\begin{equation}
\label{eq:convergence_bound}
 \frac{\norm{e_k}_A}{\norm{e_0}_A} \leq c_1(\npf) e^{-4 k \sqrt{\lambda_{\npf}/\lambda_{\mathrm{max}}}} + \mathcal{O}(k(\lambda_{\npf}/\lambda_{\mathrm{max}})^{3/2}),
\end{equation}
so we see that the rate of convergence for the block solver goes like $\sim\sqrt{\lambda_{\npf}}$, or equivalently, the effective ``condition number" that governs the convergence of the solver is reduced as $\npf$ is increased. Thus, if we keep the desired error constant, we expect the required number of iterations $k$ to decrease as we increase $\npf$, as seen in Fig.~\ref{fig:convergence}.

This solver was proposed nearly 40 years ago~\cite{OLEARY1980293}, and perhaps one reason that it has not become more widely used is its numerical stability. In particular, if the matrix of residuals $R$ becomes badly conditioned the BCG algorithm can fail to converge, while a separate CG solve for each vector for the same system would converge. Several solutions to this issue are proposed in Ref.~\cite{Dubrulle2001}, which we implemented and tested numerically, reaching the same conclusion that the optimal choice in terms of stability and computational cost is to include a re-orthogonalization via QR decomposition of the residual matrix at each iteration, known as the BCGrQ algorithm, as used in Ref.~\cite{Clark:2017ekr}.

For the RHMC we need a multi--shift variant of this solver. For CG the shift--invariance of the Krylov basis allows the residuals of the shifted systems to be related to the residuals of the unshifted one, leading to the multi--shift CG (SCG) algorithm~\cite{Frommer:1995ik,Jegerlehner:1996pm}. The same can be done for the BCGrQ algorithm, which leads to the SBCGrQ~\cite{futamura} multi--shift block solver. The main difference to the multi--shift CG solver is that in the block case the relations between shifted and unshifted systems involve $\npf\times\npf$ matrices instead of scalars.

\begin{figure}
\begin{center}
  \includegraphics[width=25em]{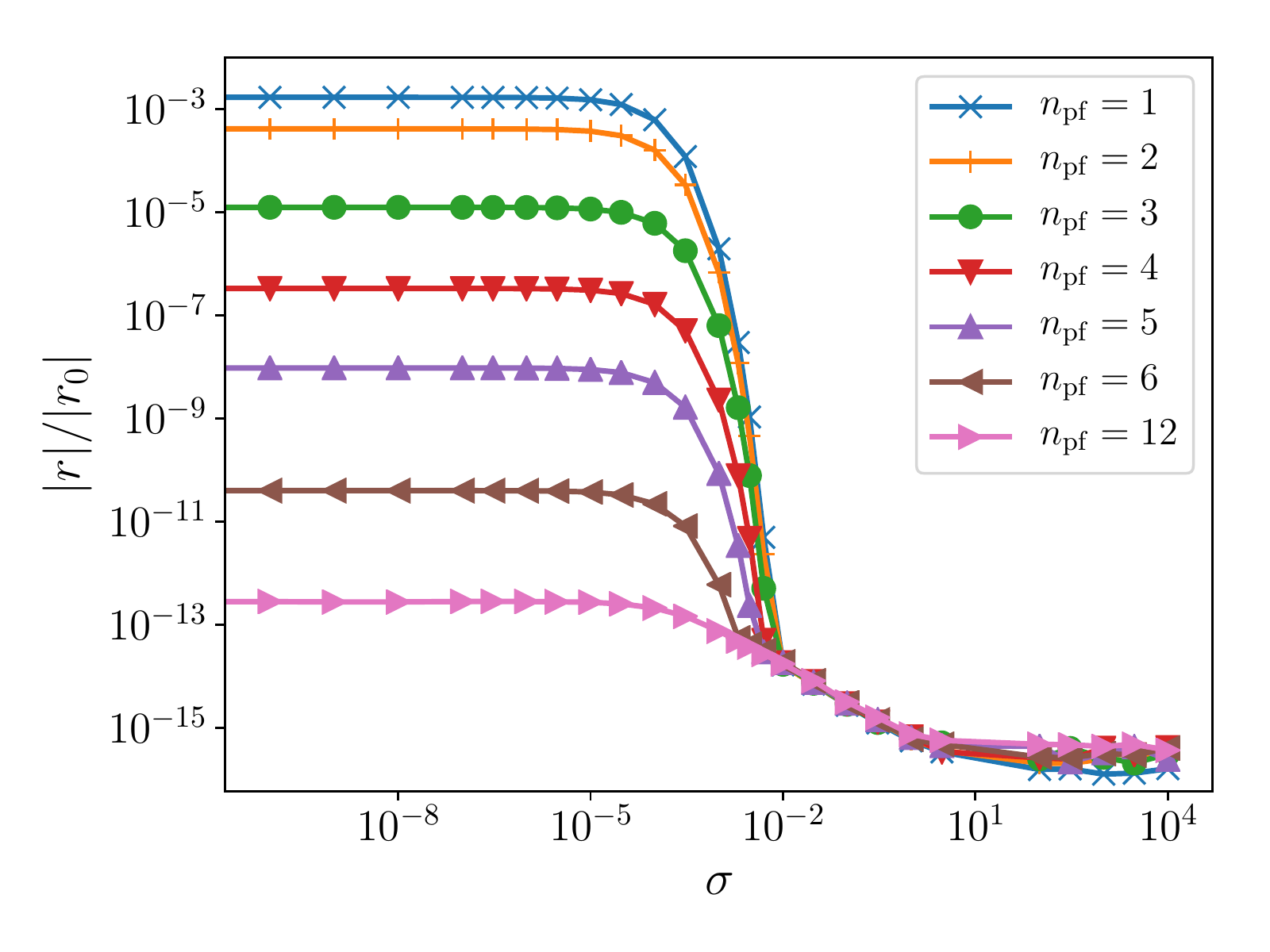}\\\includegraphics[width=25em]{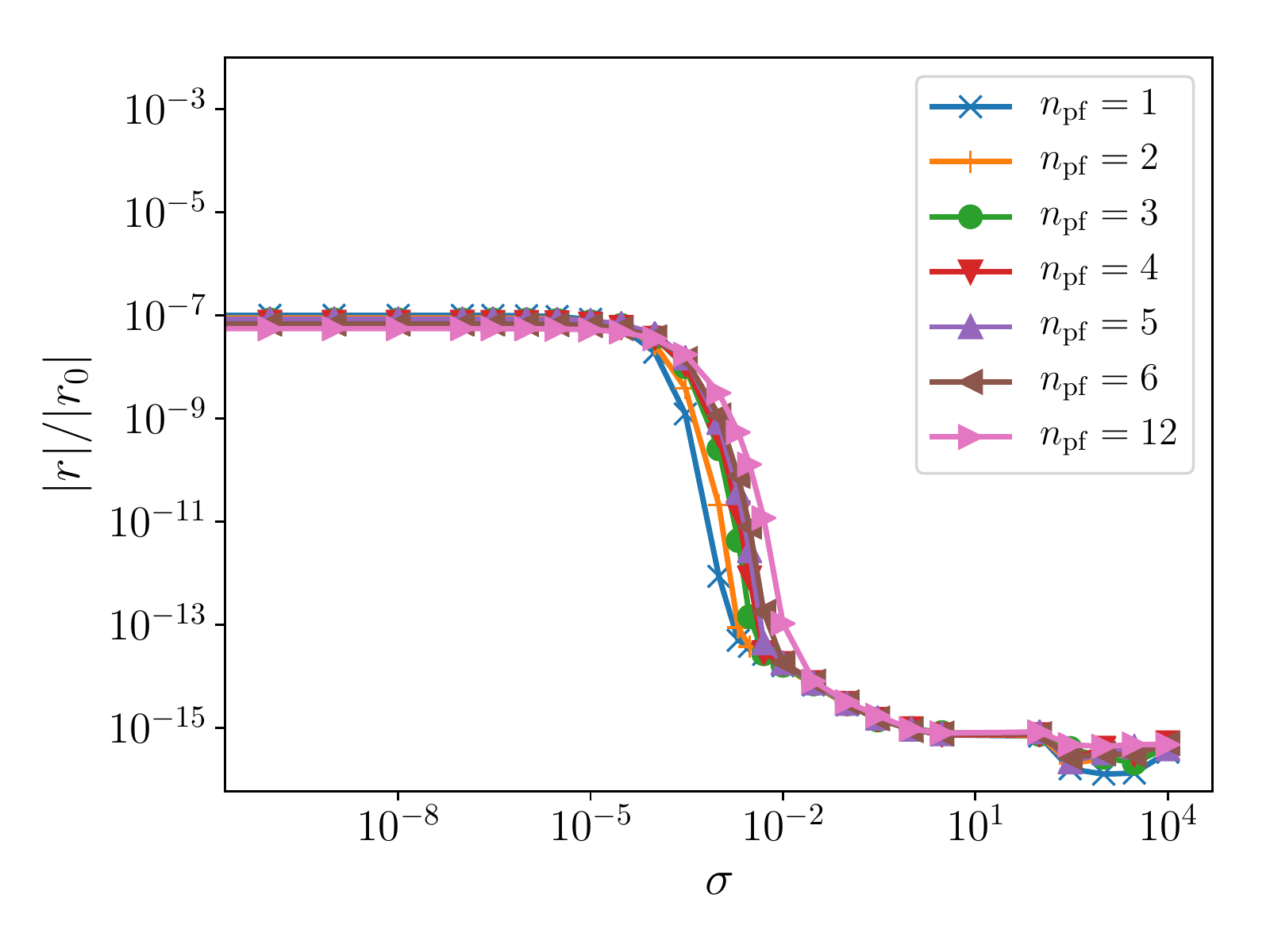}
\caption{\emph{Top}: Residual of shifted solution versus shift $\sigma$, for fixed $k=400$ solver iterations. We see a dramatic decrease in the residual for small shifts as $\npf$ is increased. \emph{Bottom}: Same quantity but keeping fixed the unshifted residual $\norm{r}/\norm{r_0}=10^{-7}$. We see that the reduction in iterations leads to a relative increase in the residuals of the larger shifts with $\npf$.}
\label{fig:residuals}
\end{center}
\end{figure}

It is instructive to consider how the bound on the error, Eq.~(\ref{eq:convergence_bound}), changes for the shifted matrix $A+\sigma$, in particular for the case where $\sigma \gg \lambda_{\npf}$,
\begin{align}
\label{eq:convergence_bound_shifts}
 \frac{\norm{e_k}_{A+\sigma}}{\norm{e_0}_{A+\sigma}} 
 &\lesssim 
 c_1(\npf) e^{-4 k \sqrt{(\sigma + \lambda_{\npf})/(\sigma + \lambda_{\mathrm{max}})}} \\ \nonumber
 &\lesssim  c_1(\npf)e^{-4 k \sqrt{\sigma/(\sigma + \lambda_{\mathrm{max}})} \left[1 + \mathcal{O}(\lambda_{\npf}/\sigma)\right]}.
\end{align}
Here we see that to leading order the convergence rate does not depend on $\lambda_{\npf}$, but only on the size of the shift $\sigma$ and the number of steps $k$. From Eq.~(\ref{eq:convergence_bound}) we expect that the number of steps $k$ required for a given error on the unshifted solution decreases with $\npf$. Eq.~(\ref{eq:convergence_bound_shifts}) suggests that, as a side-effect, the error on shifted solutions with large shifts will increase with $\npf$, as shown in Fig.~\ref{fig:residuals}.

The formulation of SBCGrQ used here is described in Algorithm~(\ref{alg:SBCGrQ}). It is numerically equivalent to Ref.~\cite{futamura}, but we use a pair of two--term coupled recursion relations instead of a single three--term recursion relation to calculate the shift matrices, which we find improves the numerical accuracy of the shifted solutions for very badly conditioned systems~\cite{doi:10.1137/S0895479897331862}. The updating of a shifted solution can be stopped once the relative norm of its residual, $\sqrt{\sum_j\delta^{(s)}_k(i,j) / \sum_j\delta_0(i,j)}$, is less than machine precision, where $\delta^{(s)}_k = \rho_{k}\alpha^{-1}_{k}\alpha^{(s)}_{k}$, see Algorithm~(\ref{alg:SBCGrQ}). A reference {\cpp} implementation of the algorithm is available at \mbox{\url{https://github.com/lkeegan/blockCG}}. Compared to BCGrQ, each shifted solution requires two additional block vectors to be stored, and two additional multiply-add operations (lines $12-13$ of Algorithm~(\ref{alg:SBCGrQ})) involving these block vectors at each iteration. There are also some extra $\npf\times\npf$ matrix operations (lines $10-11$ of Algorithm~(\ref{alg:SBCGrQ})) that have negligible storage and computational impact.
The expression $\{Q, R\} = \qr(B)$ in Algorithm~(\ref{alg:SBCGrQ}) refers to a thin QR--decomposition of the matrix $B$ into an orthogonal matrix $Q$ and an upper--triangular matrix $R$ such that $QR = B$, as described for example in Ref.~\cite{Clark:2017ekr}.

\begin{figure}
\begin{algorithm}[H]
\caption{SBCGrQ: Solve $(A+\sigma_s) X^{(s)} = B$ for $s = 0, 1, \ldots, \ns-1$}
\label{alg:SBCGrQ}
\begin{algorithmic}[1]
        \State $X^{(s)}, P^{(s)}, Q, \in \mathcal{C}^{L\times\npf}$; $\alpha, \rho, \delta, \alpha^{(s)}, \beta^{(s)} \in \mathcal{C}^{\npf\times\npf}$
        \State $X^{(s)}_0 = 0$, $\{Q_0, \delta_0\} = \qr(B), P^{(s)}_0 = Q_0$;
        \Statex $\rho_0=\delta_0, \alpha_0 = \alpha^{(s)}_0 = \beta^{(s)}_0 = 1$
        \For{$k = 1, 2, \ldots$ until $\sqrt{\sum_j\delta_k(i,j) / \sum_j\delta_0(i,j)} < \epsilon  \,\forall i$}
            \State $\alpha_k \gets (P_{k-1}^{(0)\dagger}(A+\sigma_0) P_{k-1}^{(0)})^{-1}$
            \State $\{Q_{k}, \rho_k\} \gets \qr(Q_{k-1} - (A+\sigma_0) P^{(0)}_{k-1} \alpha_k$)
            \State $X_{k}^{(0)} \gets X_{k-1}^{(0)} + P_{k-1}^{(0)} \alpha_k \delta_{k-1}$
            \State $P_{k}^{(0)} \gets Q_{k} + P_{k-1}^{(0)} \rho_{k}^{\dagger}$
            \State $\delta_{k} \gets \rho_k \delta_{k-1}$
            \For{$s = 1, \ldots, \ns-1$}
                \State $\beta^{(s)}_k \gets \left(1 + (\sigma_s - \sigma_0)\alpha_{k} + \alpha_{k}\rho_{k-1} \alpha_{k-1}^{-1} (1 - \beta^{(s)}_{k-1})\rho^{\dagger}_{k-1}\right)^{-1}$
                \State $\alpha^{(s)}_k \gets \beta^{(s)}_k \alpha_{k} \rho_{k-1} \alpha^{-1}_{k-1} \alpha^{(s)}_{k-1}$
                \State $X^{(s)}_{k} \gets X^{(s)}_{k-1} + P^{(s)}_{k-1}  \alpha^{(s)}_{k}$
                \State $P^{(s)}_{k} \gets Q_{k} + P^{(s)}_{k-1} \beta^{(s)}_{k} \rho^{\dagger}_{k}$
            \EndFor
        \EndFor
\end{algorithmic}
\end{algorithm}
\end{figure}
  
\section{Results}
\label{sec:results}
\begin{figure}
\begin{center}
  \includegraphics[width=25em]{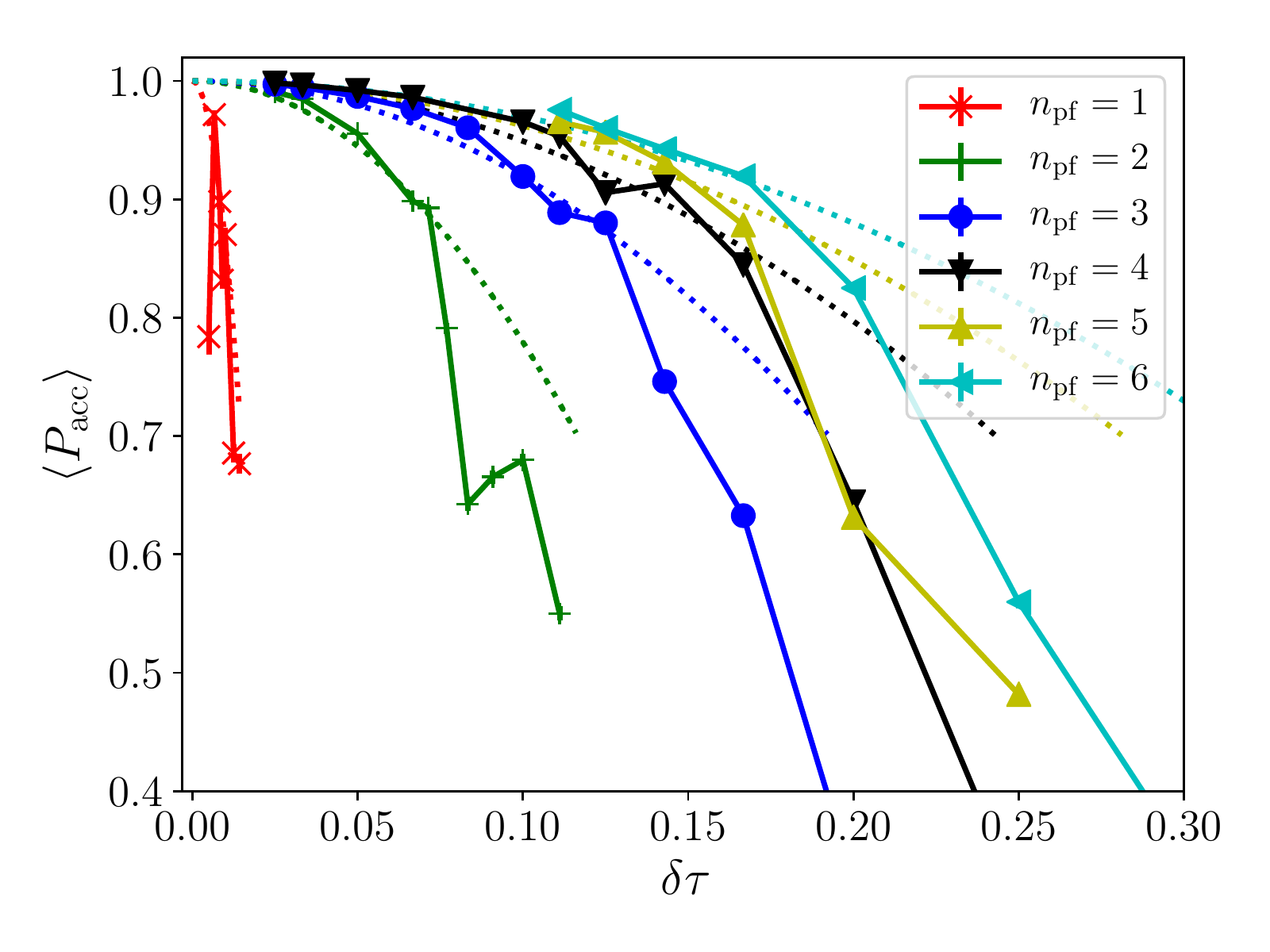}
\caption{Measured expectation values of acceptance rate (solid lines), compared with the predicted acceptance rate from measured force variances using Eq.~(\ref{eq:p_acc_force}) (dotted lines). For high acceptance and small $\delta\tau$ the agreement is reasonable; it turns out the difference between the prediction and the measured values is largely due to the neglected correlation between initial and final force terms not being negligible in these data, so increasing the trajectory length would improve the agreement.}
\label{fig:acc}
\end{center}
\end{figure}

As an initial numerical study of the method we simulate $\nf=4$ QCD using unimproved staggered fermions with even--odd preconditioning and the Wilson gauge action, on lattices of size $8^4$, with gauge coupling $\beta = 5.12$ and fermion mass $am=0.002$. These parameters are chosen to have a small mass while remaining in the confined phase of this theory~\cite{deForcrand:2017cgb}, and the choice $\nf=4$ allows a direct comparison to HMC for the case $\npf=1$ while avoiding any issues related to rooting. These small--scale simulations allow us to perform many simulations with different parameters and investigate a wide range of values of $\npf$ and integrator step sizes, as well as to perform very long simulations to study the integrated autocorrelation times of measured observables.

\begin{table}[]
\centering
\begin{tabular}{lclllcl}
\multicolumn{1}{c}{$\npf$} &
\multicolumn{1}{c}{$n_{\mathrm{steps}}$} &
\multicolumn{1}{c}{$\ex{P_{\mathrm{acc}}}$} & 
\multicolumn{1}{c}{$\ex{e^{-\Delta H}}$} & 
\multicolumn{1}{c}{$\ex{\mathrm{plaq}}$} &
\multicolumn{1}{c}{$\tau_{\mathrm{int}}$} & 
\multicolumn{1}{c}{$n_{\mathrm{trajectories}}$} \\
\hline
$1$ & $250$ & $0.961(11)$ & $0.9701(100)$ & $0.52268(14)$ & $5$ & $5\times10^3$\\
$2$ & $16$ & $0.942(5)$ & $0.9920(28)$ & $0.52283(6)$ & $4$ & $28\times10^3$\\
$3$ & $11$ & $0.965(1)$ & $0.9998(6)$ & $0.52288(8)$ & $5$ & $33\times10^3$\\
$4$ & $9$ & $0.966(1)$ & $1.0005(5)$ & $0.52297(6)$ & $4$ & $26\times10^3$\\
$5$ & $8$ & $0.960(1)$ & $0.9994(7)$ & $0.52272(8)$ & $5$ & $25\times10^3$\\
$6$ & $7$ & $0.954(2)$ & $1.0006(8)$ & $0.52277(10)$ & $6$ & $21\times10^3$\\
\end{tabular}
\caption{Run parameters for the longer simulations, with $n_{\mathrm{steps}}$ tuned such that $\ex{P_{\mathrm{acc}}}\simeq0.96$. The integrated autocorrelation time of the plaquette does not appear to depend on $\npf$.}
\label{tab:long_runs}
\end{table}

For the molecular dynamics force term we use a stopping criterion $\norm{r}/\norm{r_0} < 10^{-7}$ for the solver, and a rational approximation with relative error $< 10^{-7}$ and $\ns \simeq 15$, while for the heatbath and accept/reject steps the stopping criterion is $10^{-14}$, and the rational approximation has relative error $< 10^{-15}$ and $\ns \simeq 30$. We use a two--level OMF2 integrator, setting $\lambda=1/6$ in order to compare with the predicted acceptance rates of Eq.~(\ref{eq:p_acc_force}). For each pseudofermion integration step the gauge force is integrated with at least 3 steps, such that its contribution to the integrator error is negligible. For $\npf=1-6$ we ran 5000 $\tau=1$ trajectories for a wide range of integrator step sizes, whose acceptance rates are shown in Fig.~\ref{fig:acc}, along with the predicted acceptance rates using Eq.~(\ref{eq:p_acc_force}).
For high acceptance rates and small integrator step size $\delta\tau$, where Eq.~(\ref{eq:p_acc_force}) is valid, the measured values are in reasonable agreement with the prediction - the main source of the difference between the two in this case is the neglected contribution from the correlation between initial and final force terms in a trajectory, which is not negligible in our simulations. Increasing the trajectory length would suppress this contribution and improve the agreement between the predicted and measured acceptance rates. We also performed some additional shorter runs at larger $\npf$ up to $\npf=64$.

To study the $\npf$--dependence of the distribution of $\Delta H$ and of various observables and their autocorrelation times, we performed a single long run for each $\npf\leq 6$ as described in Table ~\ref{tab:long_runs}, using the OMF2 integrator setting $\lambda=0.20$. The expectation value of the plaquette is consistent within errors for all $\npf$. Its integrated autocorrelation time also exhibits no clear dependence on $\npf$, nor did the various other smeared and unsmeared gauge observables that we measured.

\subsection{Multiple Pseudofermions}
\label{sec:results:npf}
\begin{figure} 
\begin{center}
  \includegraphics[width=25em]{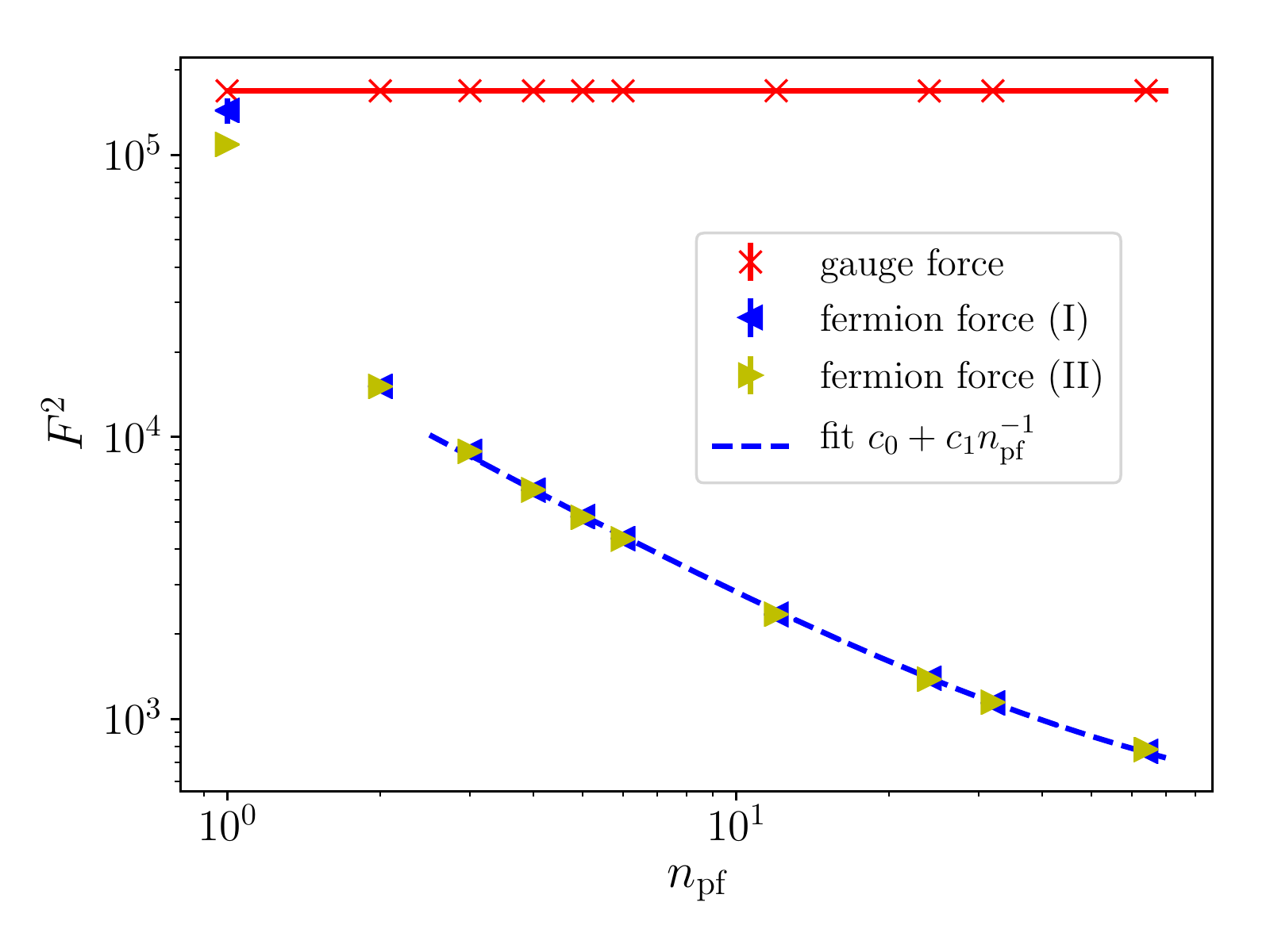}\\\includegraphics[width=25em]{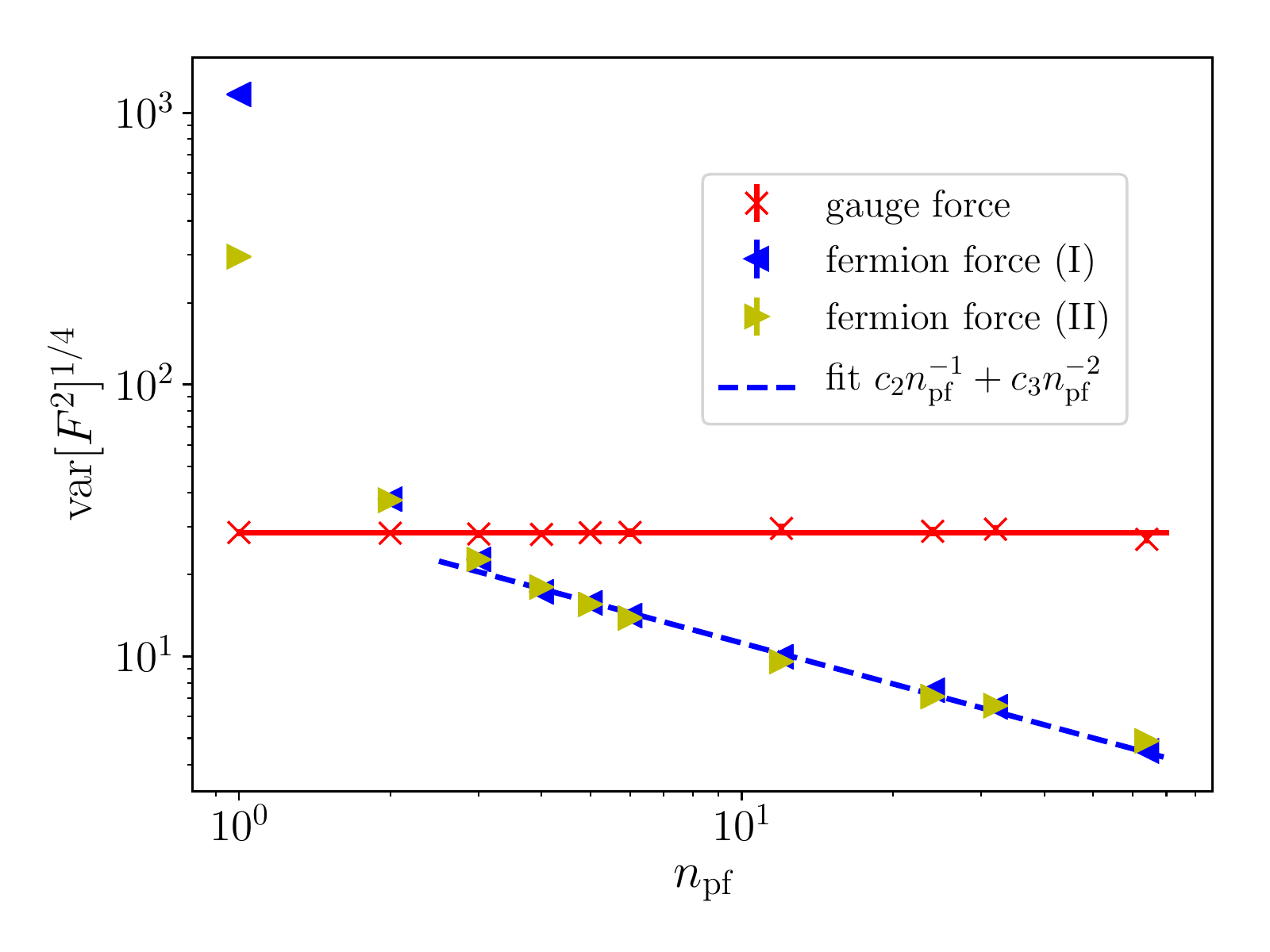}
\caption{Gauge and fermion force norms versus $\npf$, with large--$\npf$ scaling predictions. \emph{Top}: Force norms with a fit to Eq.~(\ref{eq:force_norm_npf}). \emph{Bottom}: Fourth root of variance of force norms, approximately proportional to the number of integration steps required for the OMF2 integrator, along with a fit to Eq.~(\ref{eq:force_norm_var_npf}). Force (I) is measured at every integration step along the trajectory, while Force (II) is measured on the same set of 2000 thermalised configurations.}
\label{fig:npf_F}
\end{center}
\end{figure}

\begin{figure} 
\begin{center}
  \includegraphics[width=25em]{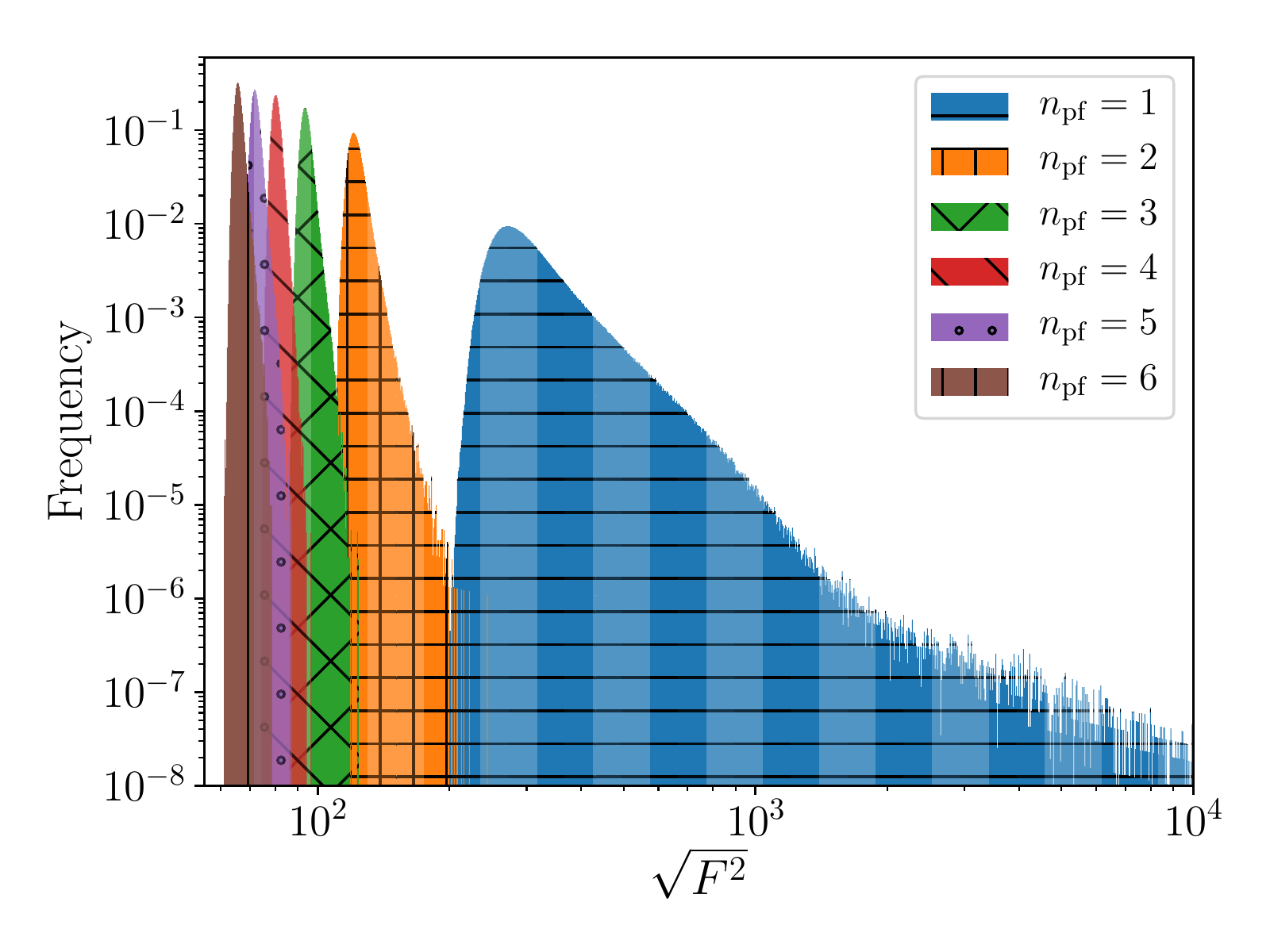}\\\includegraphics[width=25em]{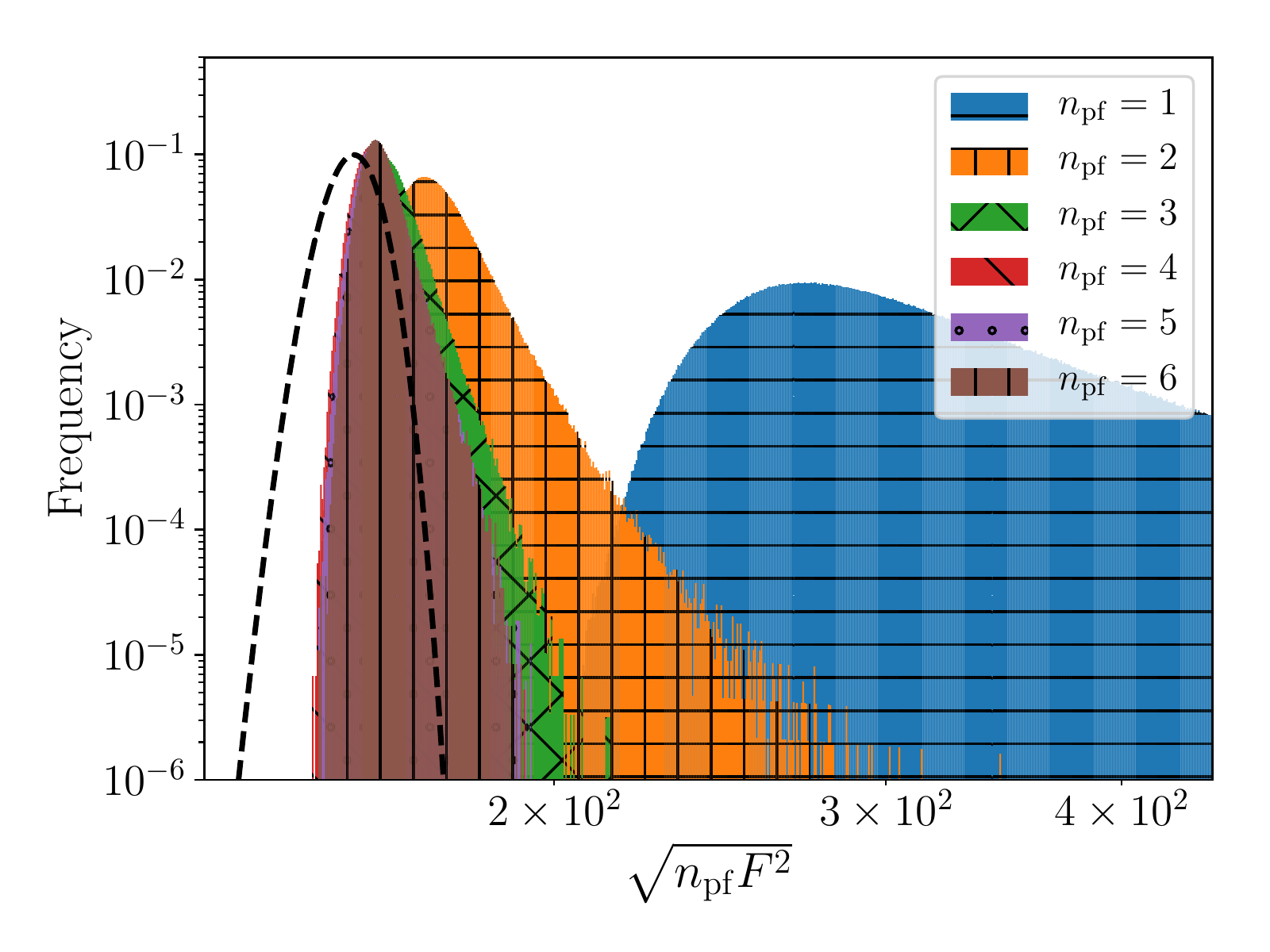}
\caption{\emph{Top}: Histogram of the rms pseudofermion force norm, $\sqrt{F^2}$, for different $\npf$. For $\npf=1$ the distribution is very non--gaussian, with a long tail of large values. \emph{Bottom}: Histogram of $\sqrt{\npf F^2}$ wich shows an approximate $\npf$--invariance for intermediate values of $\npf$, due to the $c_1$ and $c_3$ terms dominating Eqs.~(\ref{eq:force_norm_npf}, \ref{eq:force_norm_var_npf}) for these values of $\npf$.}
\label{fig:npf_F_hist}
\end{center}
\end{figure}

\begin{figure}
\begin{center}
\begin{tabular}{c c c}
   \includegraphics[width=13em]{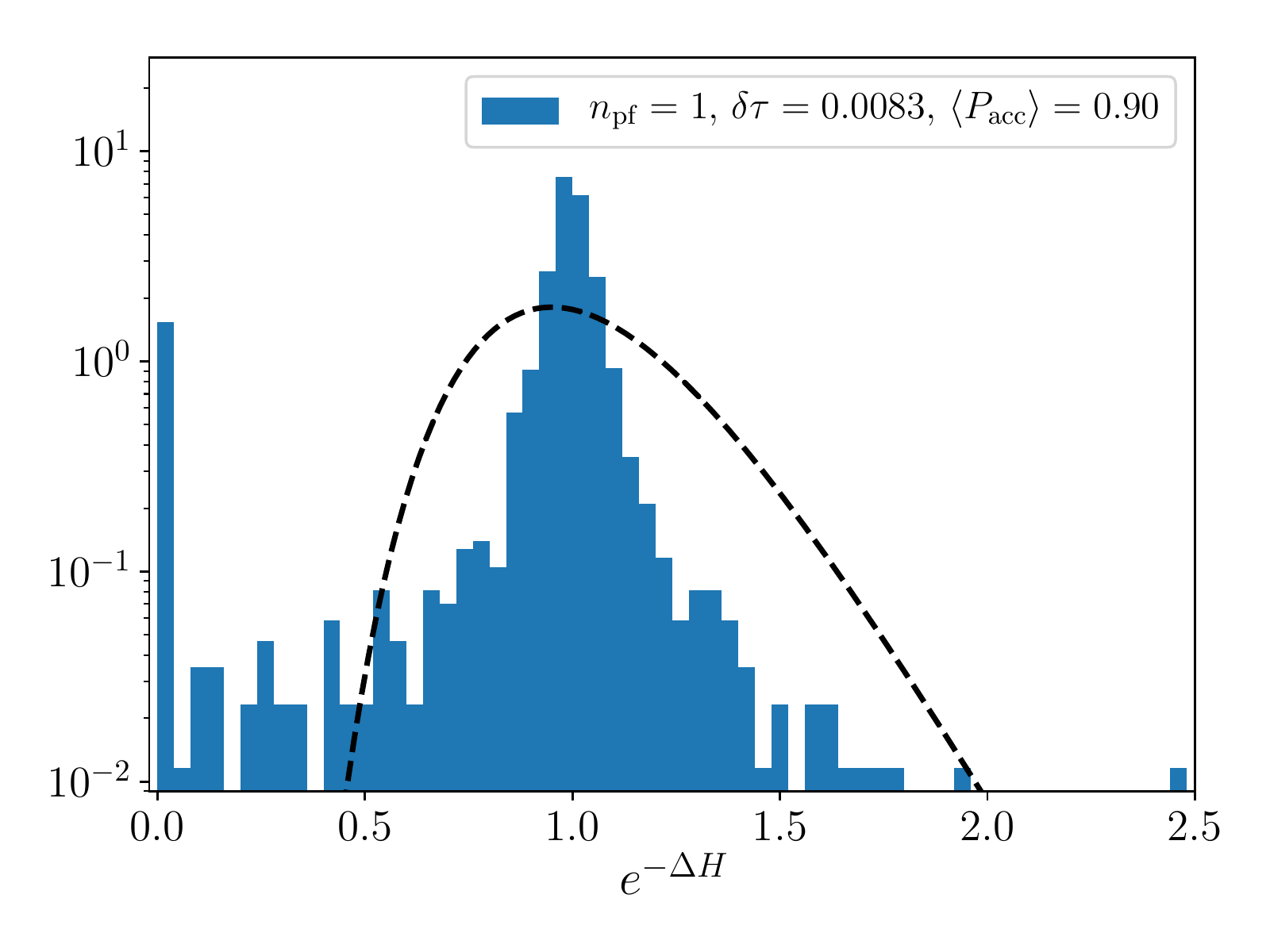} 
  &
   \includegraphics[width=13em]{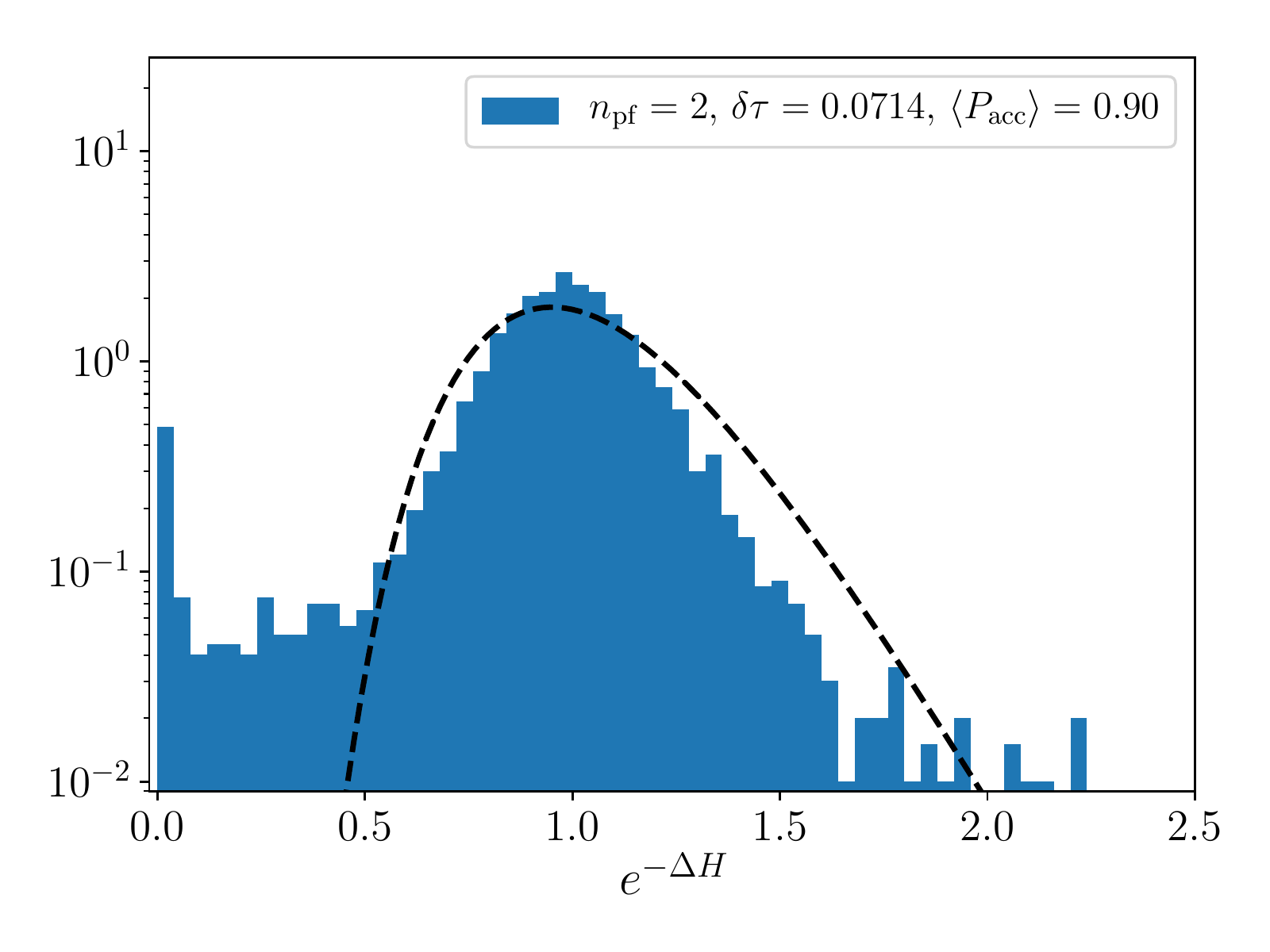} 
  \\
   \includegraphics[width=13em]{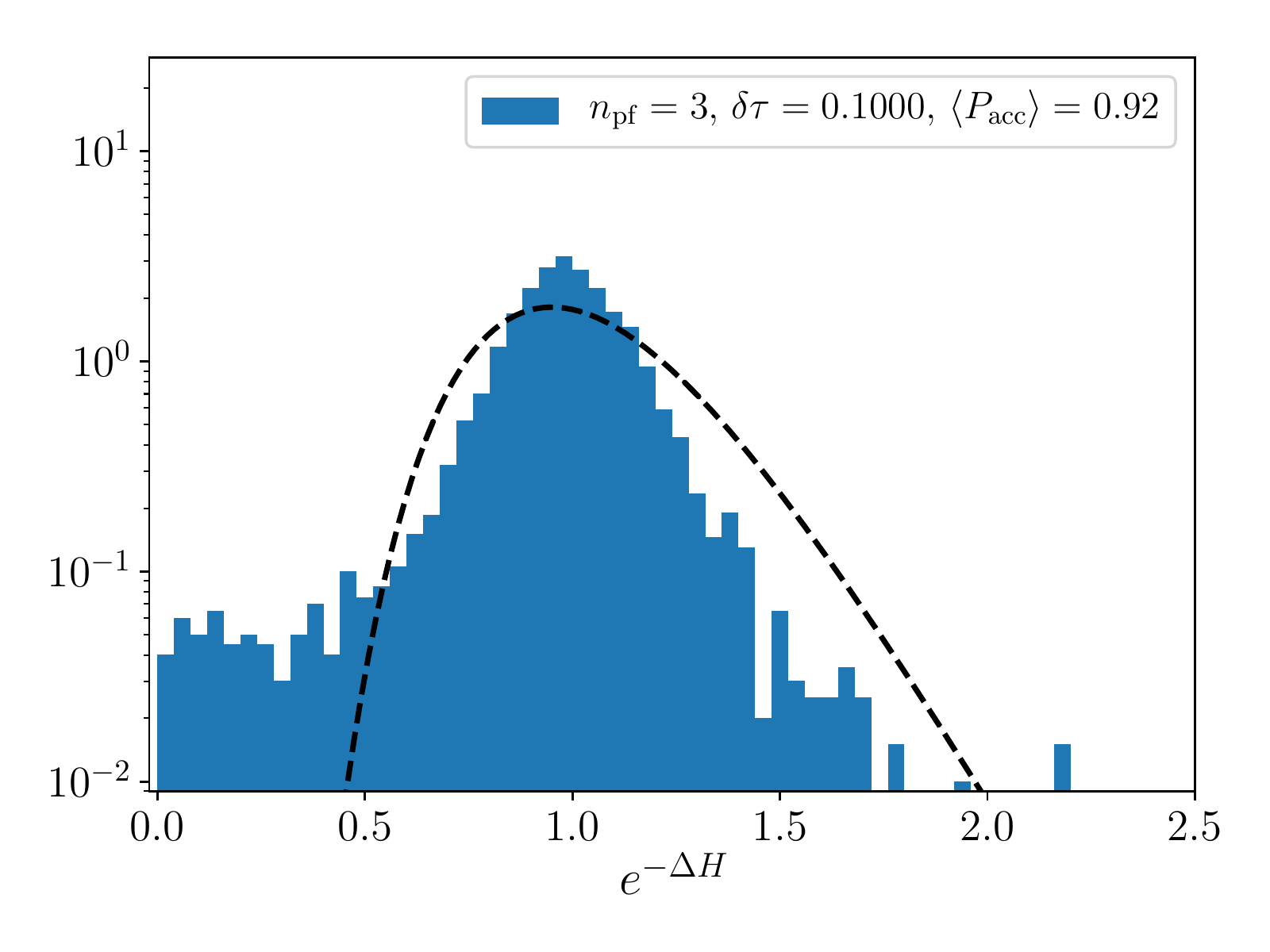} 
  &
   \includegraphics[width=13em]{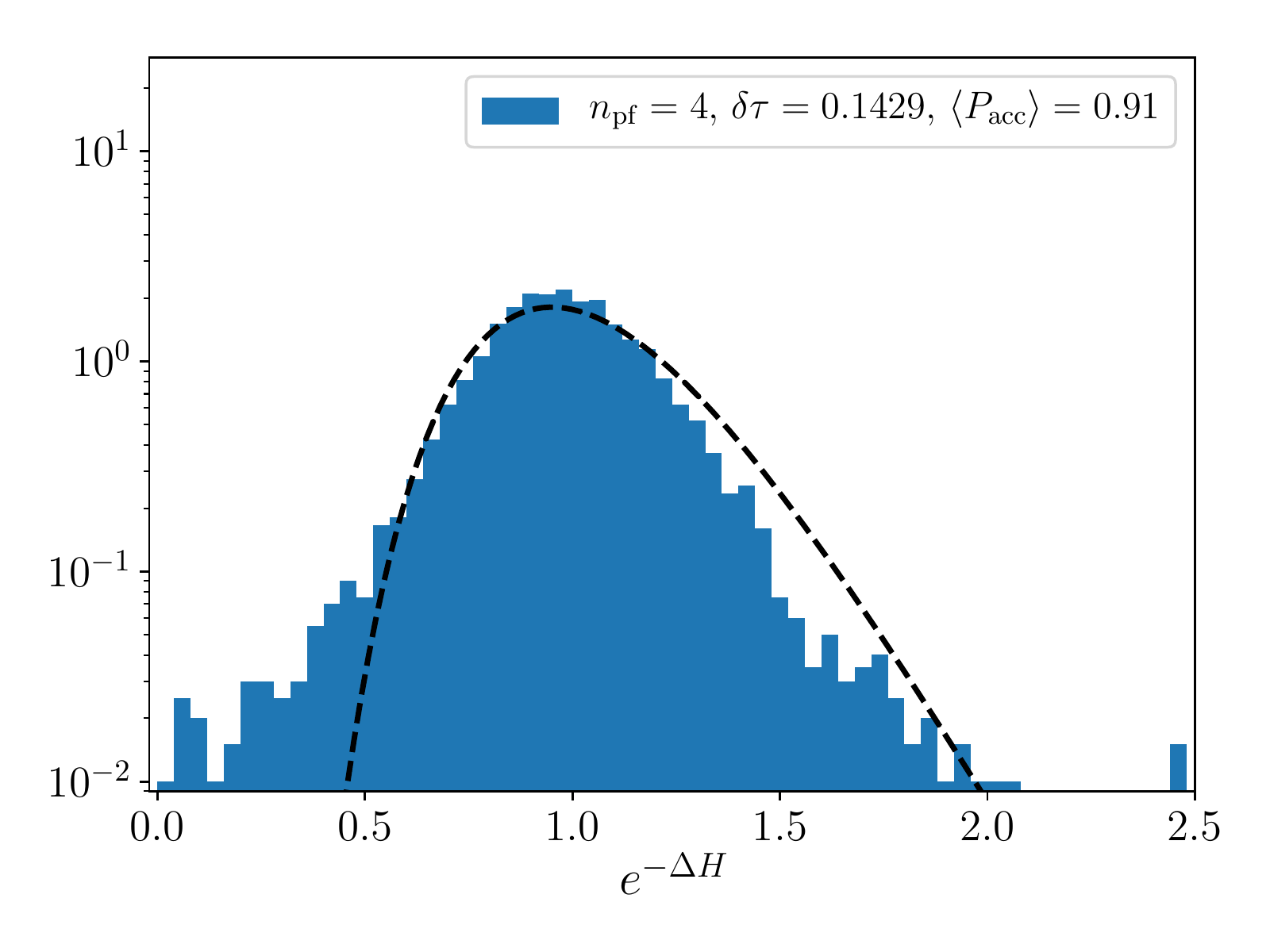} 
  \\
   \includegraphics[width=13em]{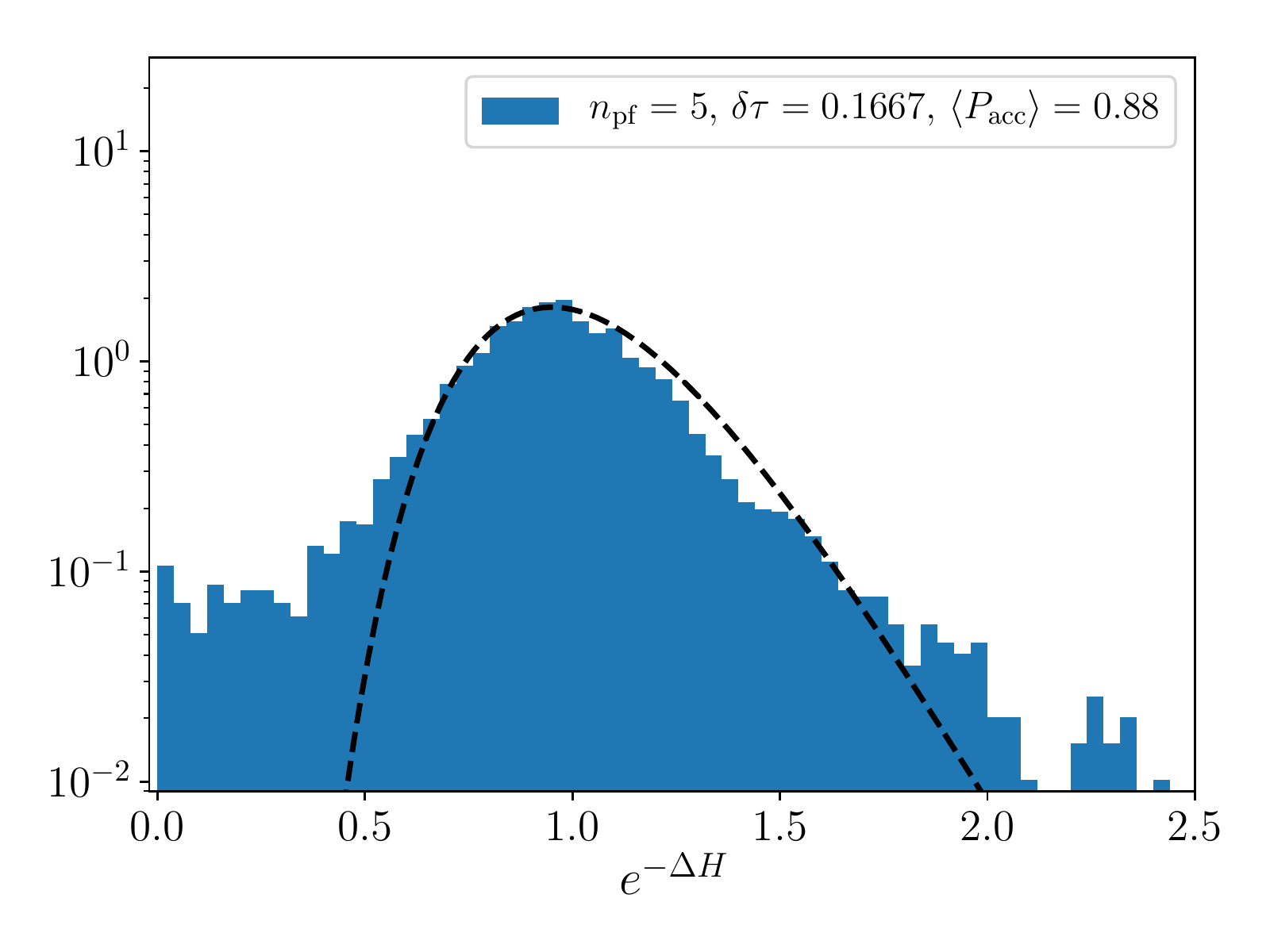} 
  &
   \includegraphics[width=13em]{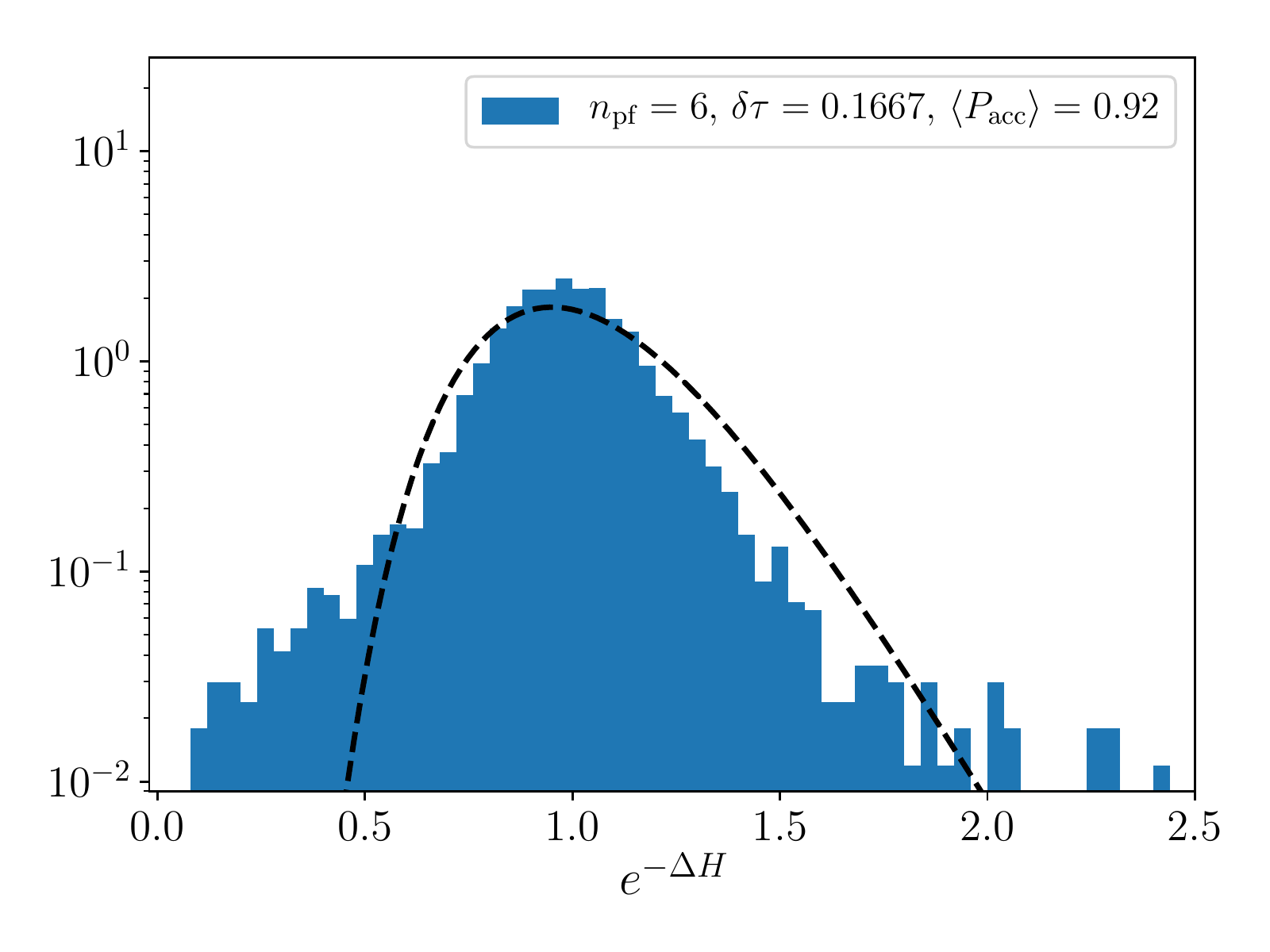} 
\end{tabular}
\caption{Histogram of $e^{-\Delta H}$ for $\npf=1$ to $6$ with the acceptance rate tuned to $\simeq 90\%$. The black dotted line shows the prediction for a gaussian distribution of $\Delta H$ with the same acceptance rate. For $\npf=1$ (top left) the distribution is very far from gaussian, with an excess of very small values, but as $\npf$ is increased the distribution approaches the gaussian one.} 
\label{fig:npf_dH}
\end{center}
\end{figure}

\begin{figure}
\begin{center}
  \includegraphics[width=25em]{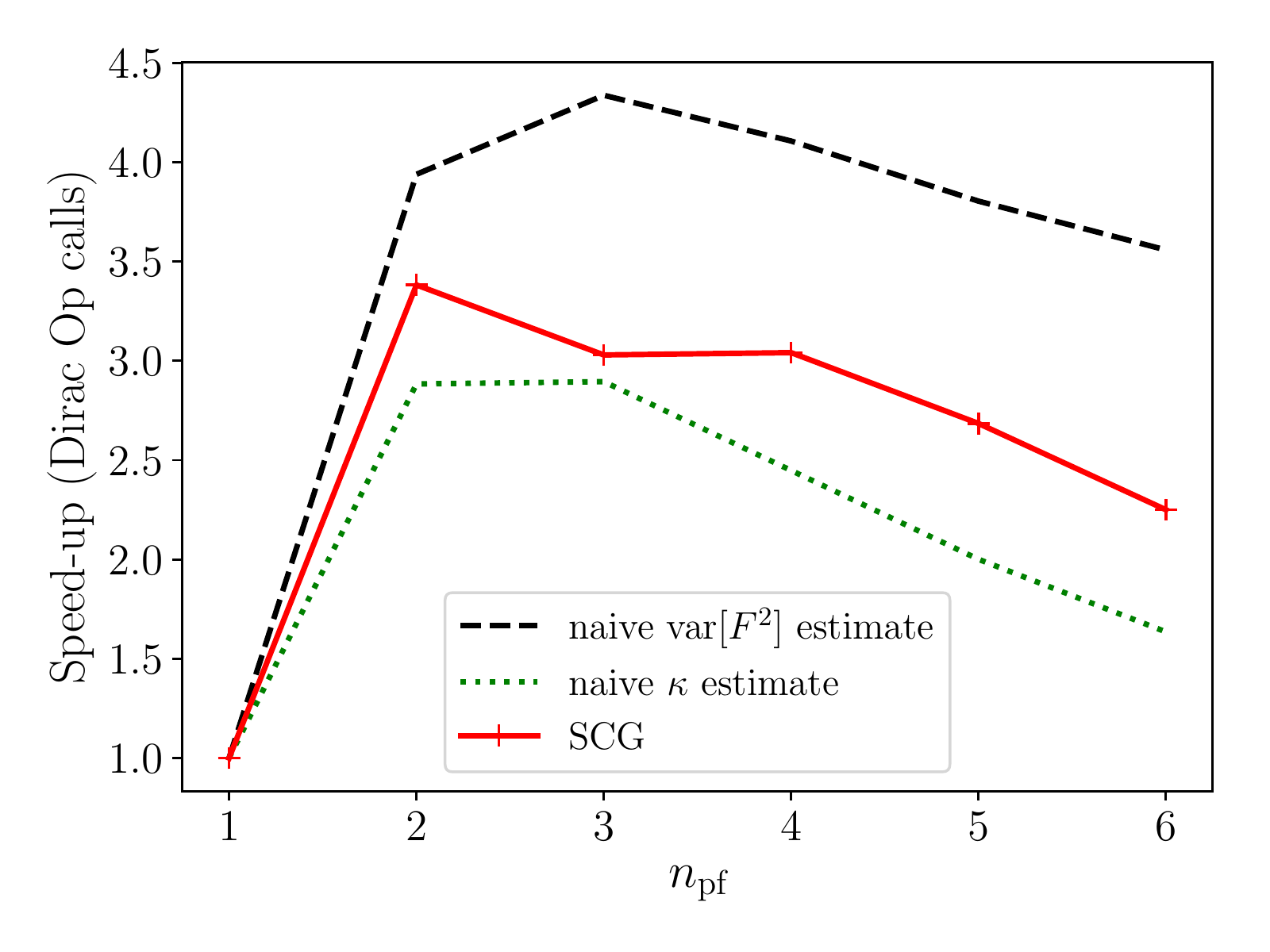}\\\includegraphics[width=25em]{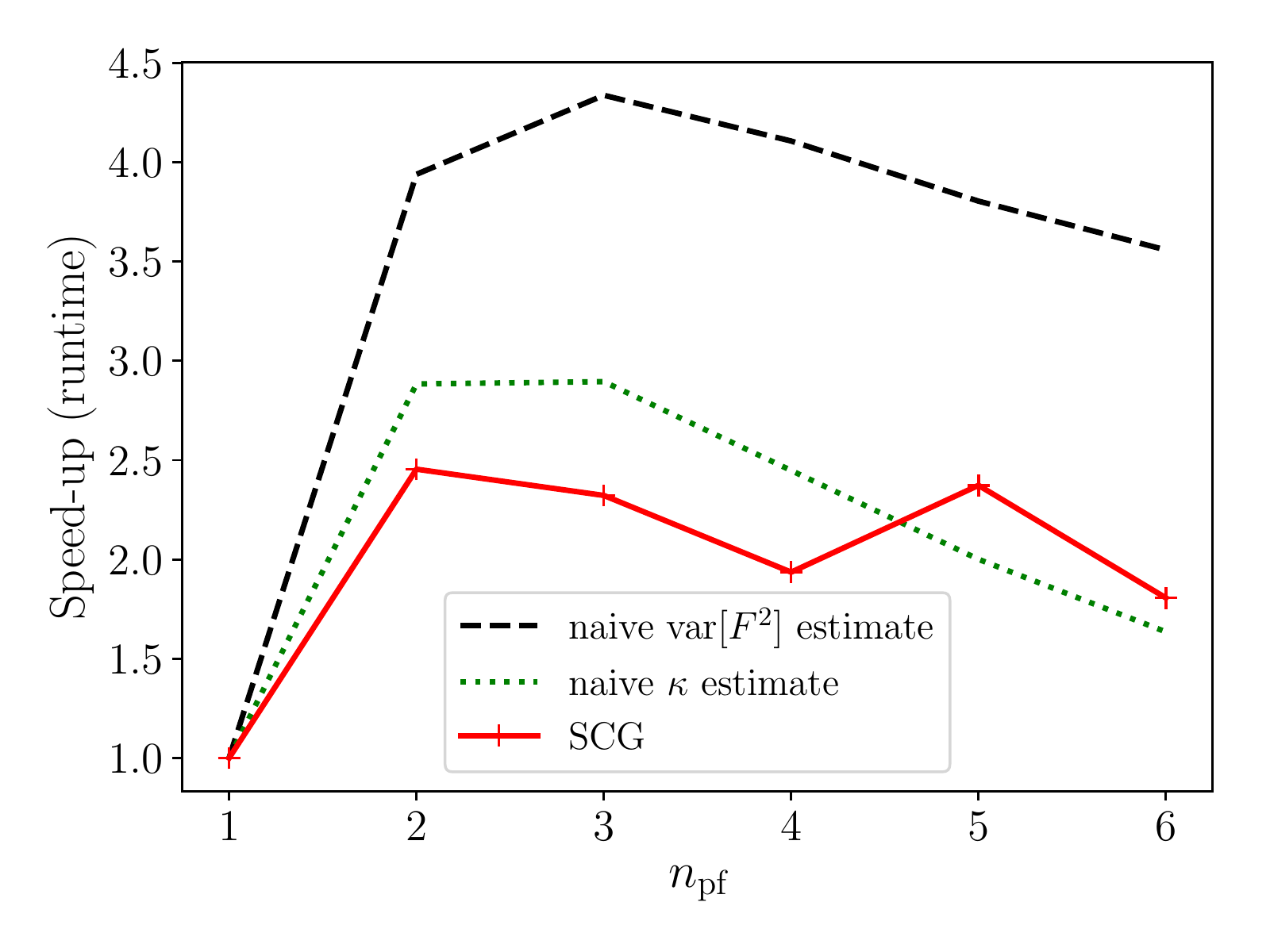}
\caption{Trajectory speed-up versus $\npf$, normalised to 1 for $\npf=1$. At each step the multishift CG (SCG) solver is used $\npf$ times. The black dashed line is the simple prediction from the variance of the pseudofermion force norm using Eq.~($\ref{eq:cost}$), and the green dotted line is the simple prediction from the condition number of the Dirac operator using Eq.~(\ref{eq:cost_clark}). Going from $\npf=1$ to $\npf=2$ gives a significant cost reduction, but increasing $\npf$ further results in a larger cost per trajectory.}
\label{fig:cost_cg}
\end{center}
\end{figure}

Increasing $\npf$ reduces both the size and the variance of the norm of the pseudofermion force term. Fig.~\ref{fig:npf_F} shows these quantities for both gauge and pseudofermion fields as a function of $\npf$. The large variance of the fermionic force comes from the poor accuracy of this pseudofermion estimate - for small $\npf$ it is orders of magnitude larger than the exact (large--$\npf$ limit) value: $c_0/c_1 \sim 10^{-3}$ in Eq.~(\ref{eq:force_norm_npf}). The blue left-facing triangles with error bars are measured for every force term calculation during the simulation, while the yellow right-facing triangles with error bars are measured on a set of 2000 thermalised configurations. For $\npf>1$, the two measurements agree within errors, but for $\npf=1$ they differ significantly. This is caused by infrequent but very large spikes in the force for $\npf=1$, which means that many more than 2000 measurements would be required to reliably estimate the variance of the force in this case. Also shown is a fit to the large--$\npf$ form predicted by Eqs.~(\ref{eq:force_norm_npf}, \ref{eq:force_norm_var_npf}), which seems to provide a good description of the data for $\npf \gtrsim 3$.

A histogram of the values of the pseudofermion rms force is shown in the top panel of Fig.~\ref{fig:npf_F_hist}, where for $\npf=1$ the distribution is clearly non--gaussian, with a long tail of large values. As $\npf$ is increased, the mean and variance of the distribution of force norms decrease, as already seen in Fig.~\ref{fig:npf_F}, and in addition the form of the distribution becomes closer to a gaussian, without a long tail of values much larger than the mean. Since empirically we find $c_0 \ll c_1 $ and $c_2 \ll c_3$ in Eqs.~(\ref{eq:force_norm_npf}, \ref{eq:force_norm_var_npf}), we can expect the quantity $\npf F^2(\npf)$ to have approximately $\npf$--independent mean and variance for some intermediate range of values of $\npf$. This quantity is shown in the bottom panel of Fig.~\ref{fig:npf_F_hist}, which shows this approximate scaling for intermediate $\npf$, along with a dotted black line showing a gaussian distribution with the same mean and variance.

Another way to see the improvement from using multiple pseudofermions is to look at the distribution of $e^{-\Delta H}$, where $\Delta H$ is the energy violation of the trajectory. Fig.~\ref{fig:npf_dH} shows the distribution of this quantity for $\npf=1$ to $6$, with the integrator step size tuned such that the acceptance is $\simeq 90\%$ for each. The distribution expected for this acceptance rate assuming a gaussian distribution for $\Delta H$ is also shown, and as $\npf$ is increased the measured distribution becomes closer to the gaussian one. For the case $\npf=1$, the distribution of $\Delta H$ is very far from gaussian, with an excess of tiny values of $e^{-\Delta H}$ which reflect the large fluctuations in the force term. Such ``exceptional configurations'' can trigger an instability of the integrator, which makes the Monte Carlo error analysis more delicate and may introduce long autocorrelation times.

Using Eq.~(\ref{eq:cost}) we can use the variance of the pseudofermion force norm to predict the approximate cost of generating an RHMC trajectory as a function of $\npf$. Another prediction of the cost using the condition number of the Dirac operator is given by Eq.~(\ref{eq:cost_clark}). These predictions are compared to the measured cost of actual simulations using the multishift CG solver, with the integrator step size tuned to make the acceptance rate $\simeq90\%$. The results are shown in Fig.~\ref{fig:cost_cg}, where all costs are normalised to 1 for the case $\npf=1$. There is a large reduction in the cost for $\npf=2$ compared to $\npf=1$, followed by a gradual increase in the cost with $\npf$.

In this section we have shown that using multiple pseudofermions with the usual multishift CG solver significantly reduces the mean and variance of the pseudofermion force term, which both speeds up RHMC simulations and results in a much more gaussian distribution of $\Delta H$. In the next section we take advantage of having multiple pseudofermions to store them in block form, which allows us to make use of a more efficient, block version of the multishift CG solver and also increases the computational efficiency of the Dirac operator.

\subsection{Block Solvers}
\label{sec:results:block}

Block solvers have been shown to provide large speed-ups in two recent lattice QCD studies of inverting the Dirac operator with multiple right hand side (RHS) vectors~\cite{Nakamura:2011my,Clark:2017ekr}. There are two sources of this speed-up: one is that as the number of RHS vectors ($\npf$ in our case) is increased the number of iterations required for the solver to converge decreases, the other is that applying the Dirac operator to a block of vectors is significantly faster, since the cost of loading the gauge links is amortised over the many RHS vectors, and these data are contiguous allowing better use of the CPU cache.

\begin{figure}
\begin{center}
  \includegraphics[width=25em]{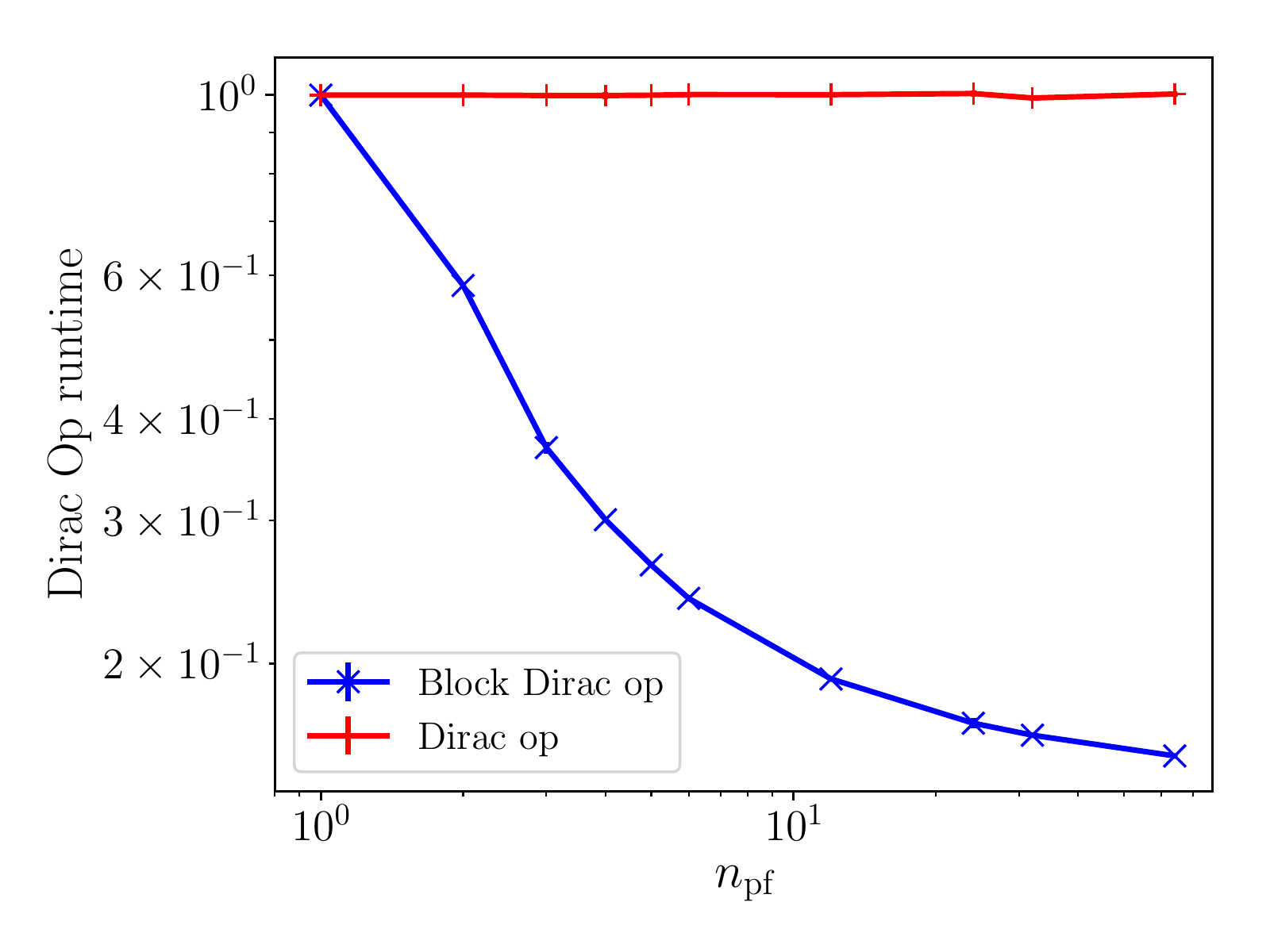}\\\includegraphics[width=25em]{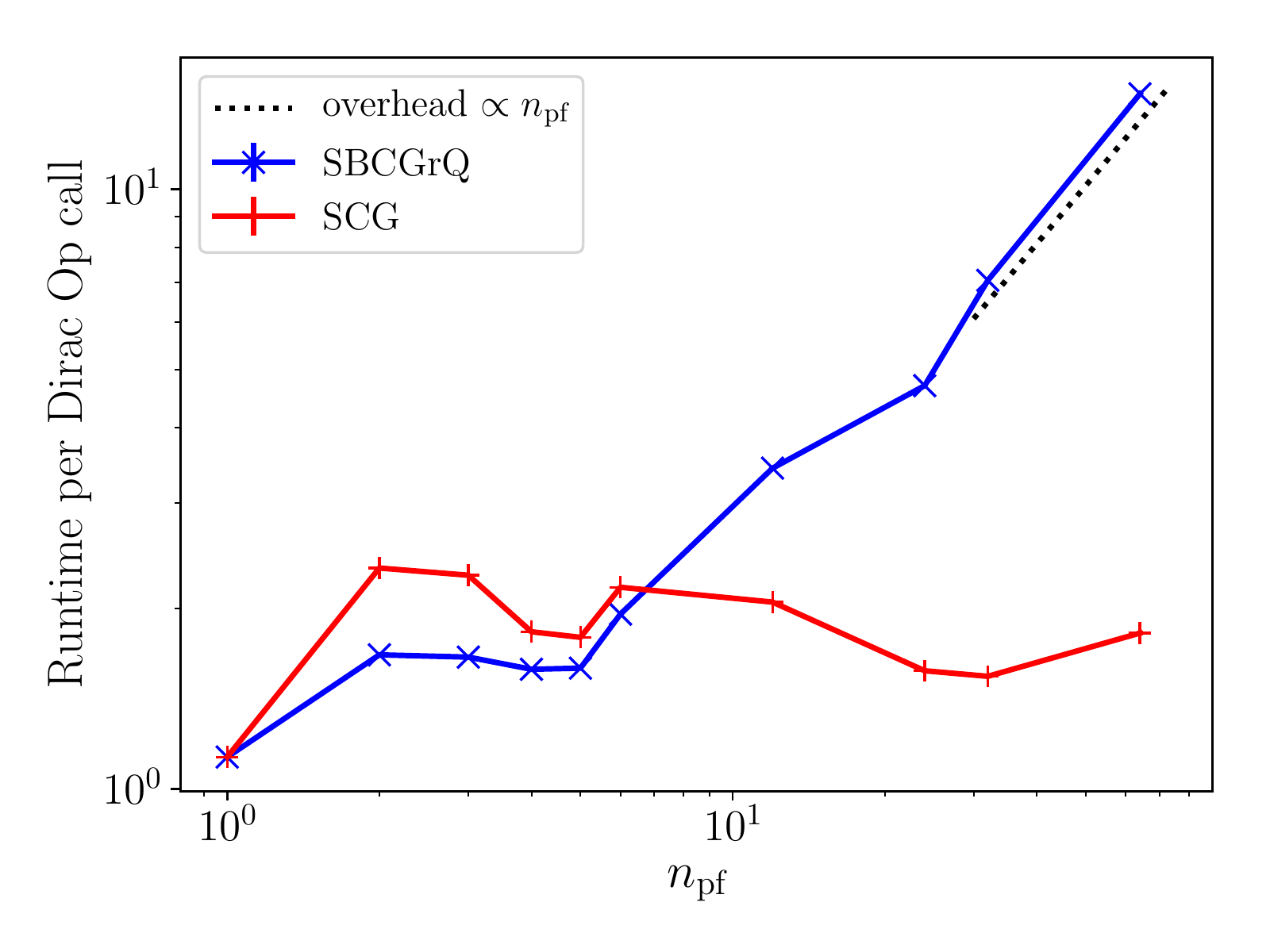}
\caption{\emph{Top}: Runtime of Dirac operator acting on vectors in block form, normalised to the non--block form. \emph{Bottom}: Solver runtime per Dirac operator call versus $\npf$. For small $\npf$ one iteration of block multishift SBCGrQ is much faster than multishift SCG since the block Dirac operator is faster. For large enough $\npf$ however, the SBCGrQ solver overhead that grows $\propto\npf \ns$ eventually dominates the cost.}
\label{fig:solver_overhead}
\end{center}
\end{figure}

However, there is a cost that comes with these benefits, which is that all pseudofermion vector operations in the solver are promoted to matrix operations in the block solver, and this overhead grows with a factor $\npf$ compared to the cost of applying the Dirac operator.
Fig.~\ref{fig:solver_overhead} compares the runtime of block and non--block versions of a single Dirac operator call and a single iteration of the two multishift solvers used in this work: multishift CG (SCG) and block multishift CG (SBCGrQ). The top panel shows that the block Dirac operator is significantly faster than the non-block version. In the bottom panel, for $\npf\leq6$ one iteration of the block multishift solver SBCGrQ is also faster than multishift CG for the same reason, because the cost is dominated by the Dirac operator. For very large $\npf$ the overhead becomes significant however, and can be seen to dominate the cost of a single SBCGrQ iteration for $\npf\gtrsim20$.

\begin{figure}
\begin{center}
  \includegraphics[width=25em]{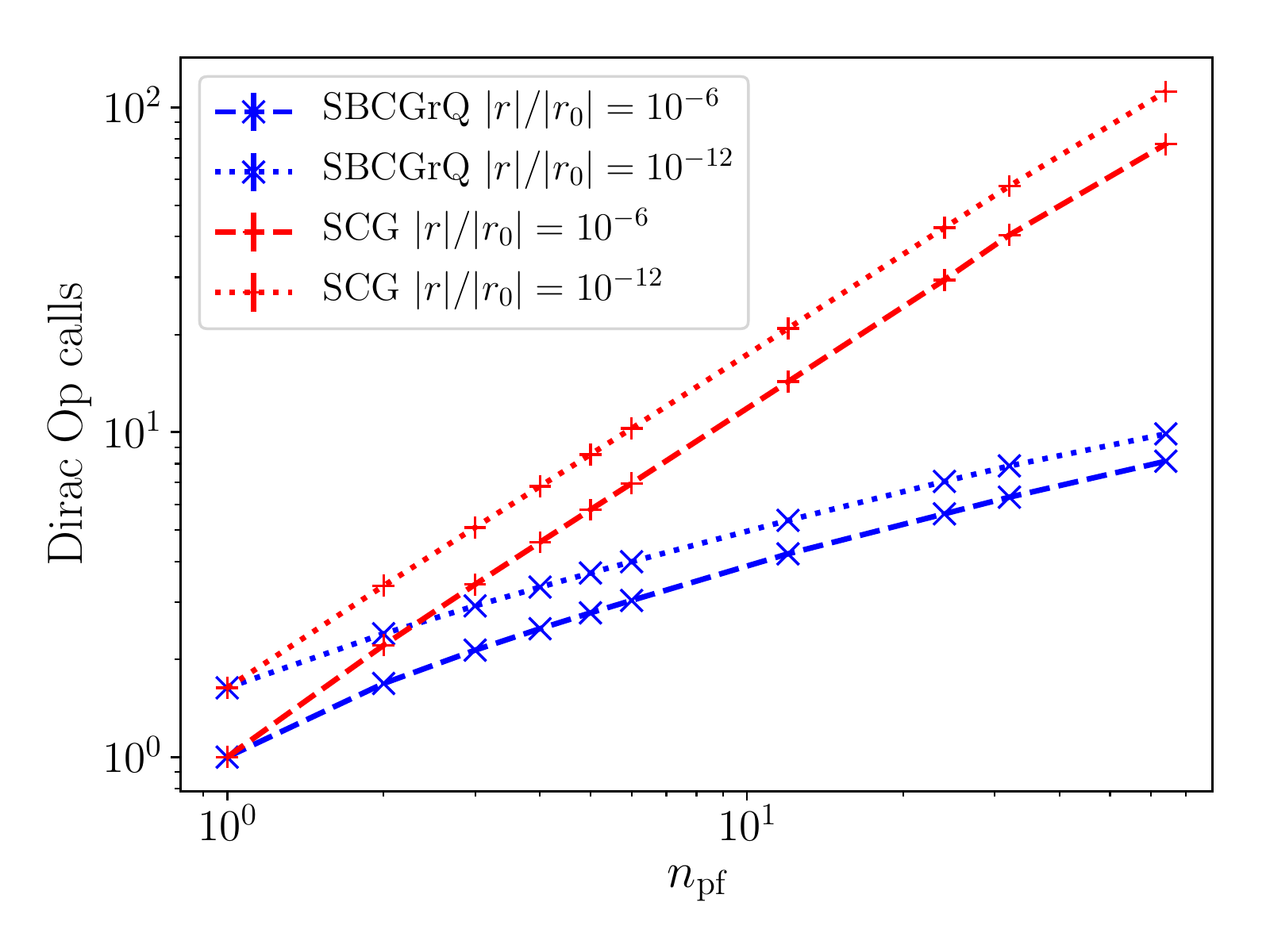}\\\includegraphics[width=25em]{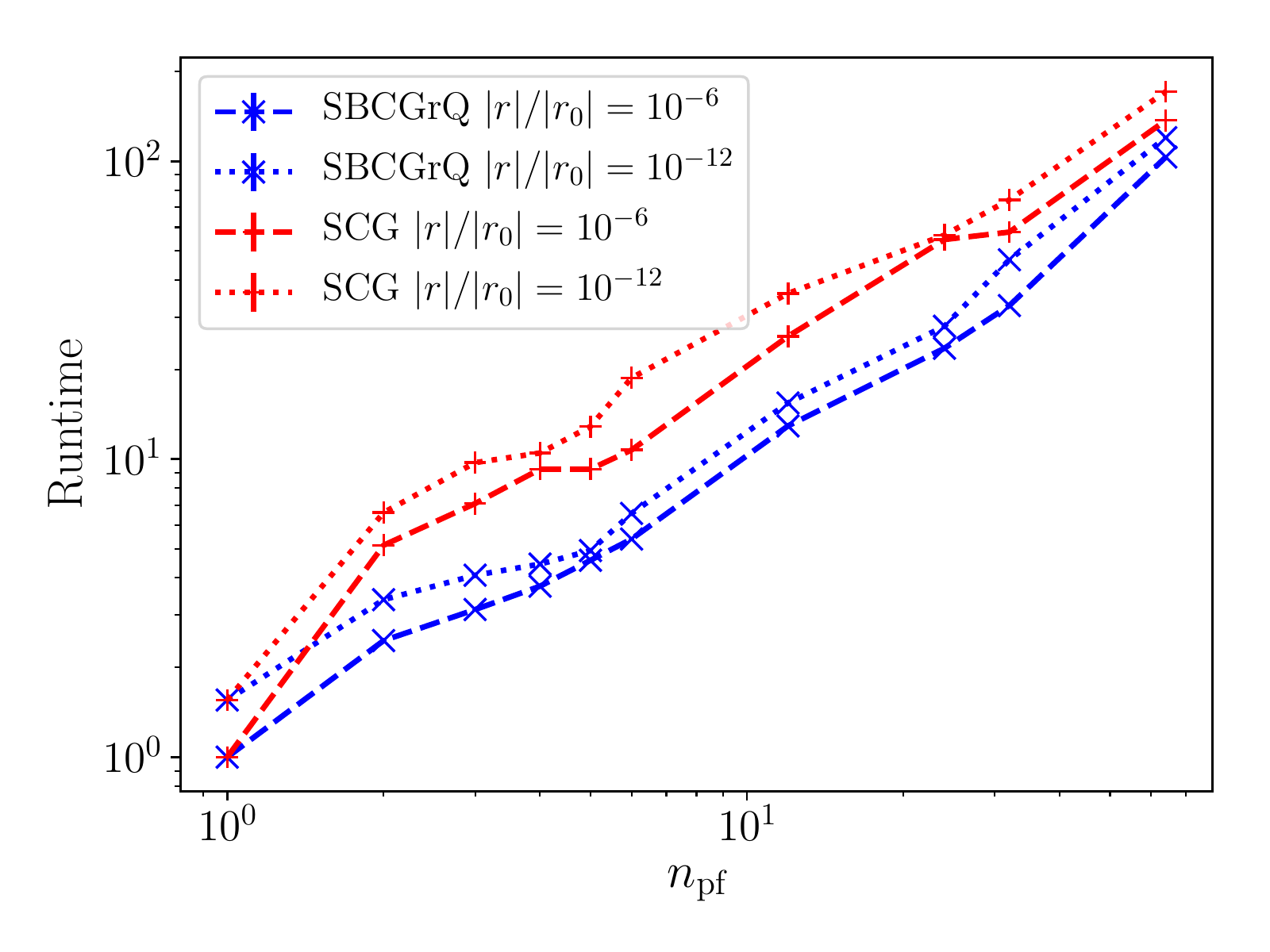}
\caption{Cost of calculating the pseudofermion force term versus $\npf$, using either multishift (SCG) or block multishift (SBCGrQ) solvers with stopping criterion $10^{-6}$ or $10^{-12}$. The block solver is a significant improvement, moreover it allows the use of a very tight stopping criterion without significant extra cost, which reduces possible reversibility violations.}
\label{fig:force_cost}
\end{center}
\end{figure}

Fig.~\ref{fig:force_cost} compares the cost of calculating the pseudofermion force term using the block multishift CG (SBCGrQ) solver with pseudofermions in block form against the previous results using the multishift CG (SCG) solver. We see a large reduction in both the number of Dirac operator calls and the overall runtime. The overhead of the SBCGrQ algorithm will eventually dominate the cost at large $\npf$, but as we already saw in Fig.~\ref{fig:solver_overhead},  for the region of interest, $\npf\lesssim6$, this overhead is not prohibitive.
It is also possible when using the block solver to take the stopping criterion for the force solves to be very small without a significant increase in cost, which reduces the potential reversibility violations caused by finite precision, which may be a concern for badly conditioned systems or if the RHMC trajectory length $\tau$ is increased~\cite{Meyer:2006ty}.

\begin{figure}
\begin{center}
  \includegraphics[width=25em]{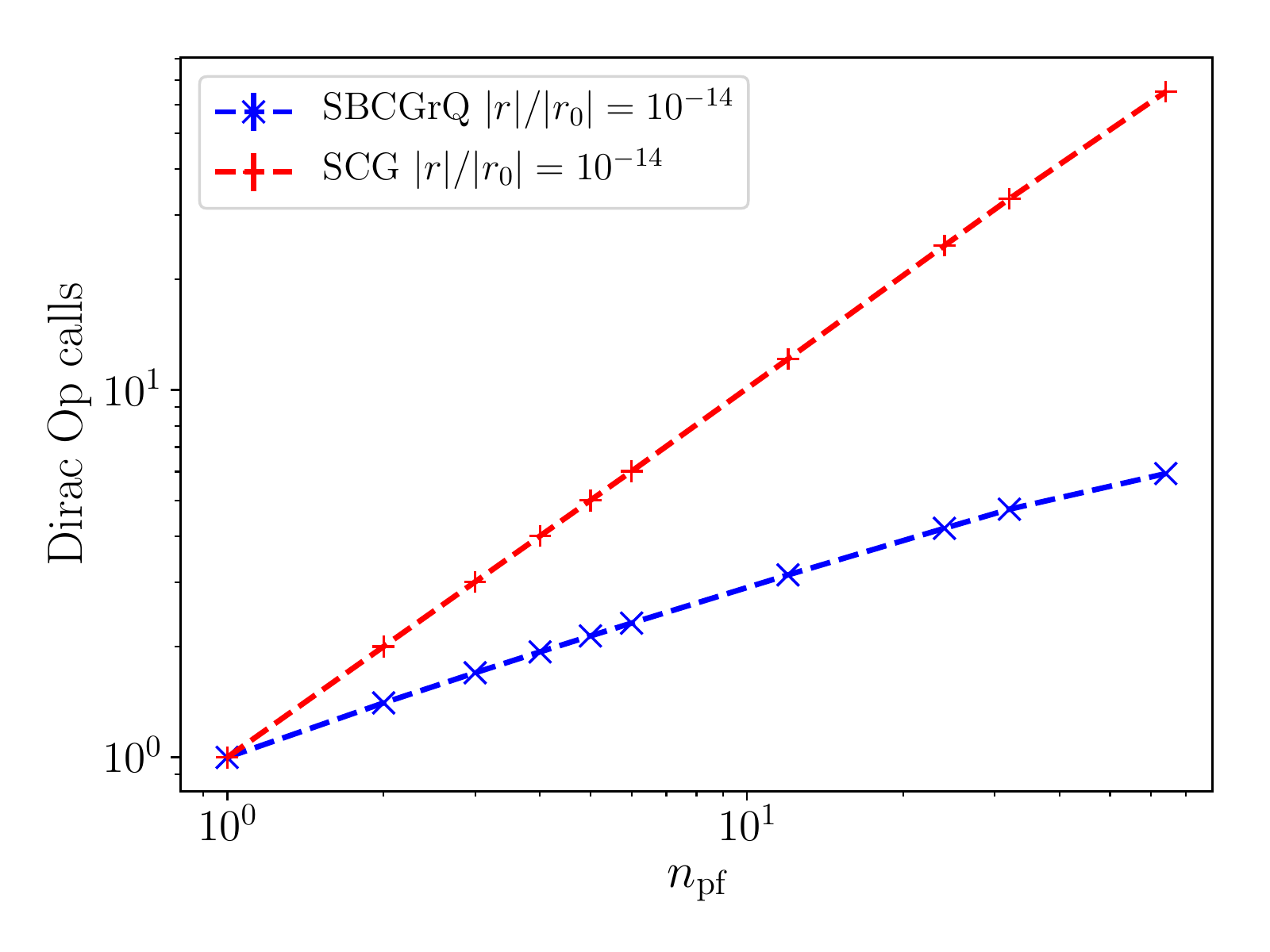}\\\includegraphics[width=25em]{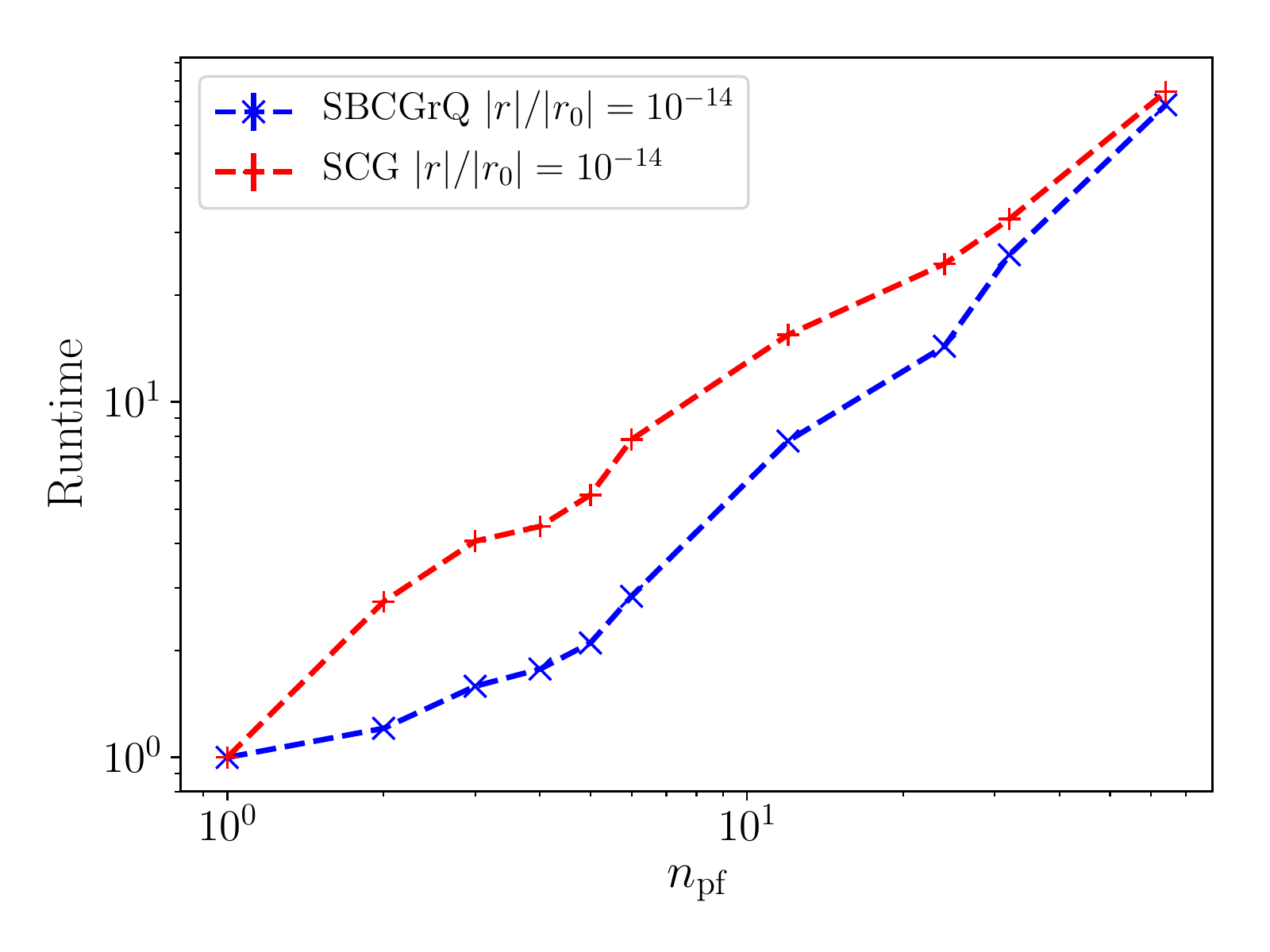}
\caption{Cost of calculating the pseudofermion action versus $\npf$, using either multishift (SCG) or block multishift (SBCGrQ) solvers with stopping criterion $10^{-14}$. This is done twice per trajectory: at the start for the heatbath and at the end for the accept/reject step. The block solver significantly reduces the cost of this step.}
\label{fig:hb_cost}
\end{center}
\end{figure}

At the start and end of a trajectory, a high precision inversion must also be done, and Fig.~\ref{fig:hb_cost} compares the cost of this step between the original and block method, and we again see a large improvement from the block version.

\begin{figure}
\begin{center}
  \includegraphics[width=25em]{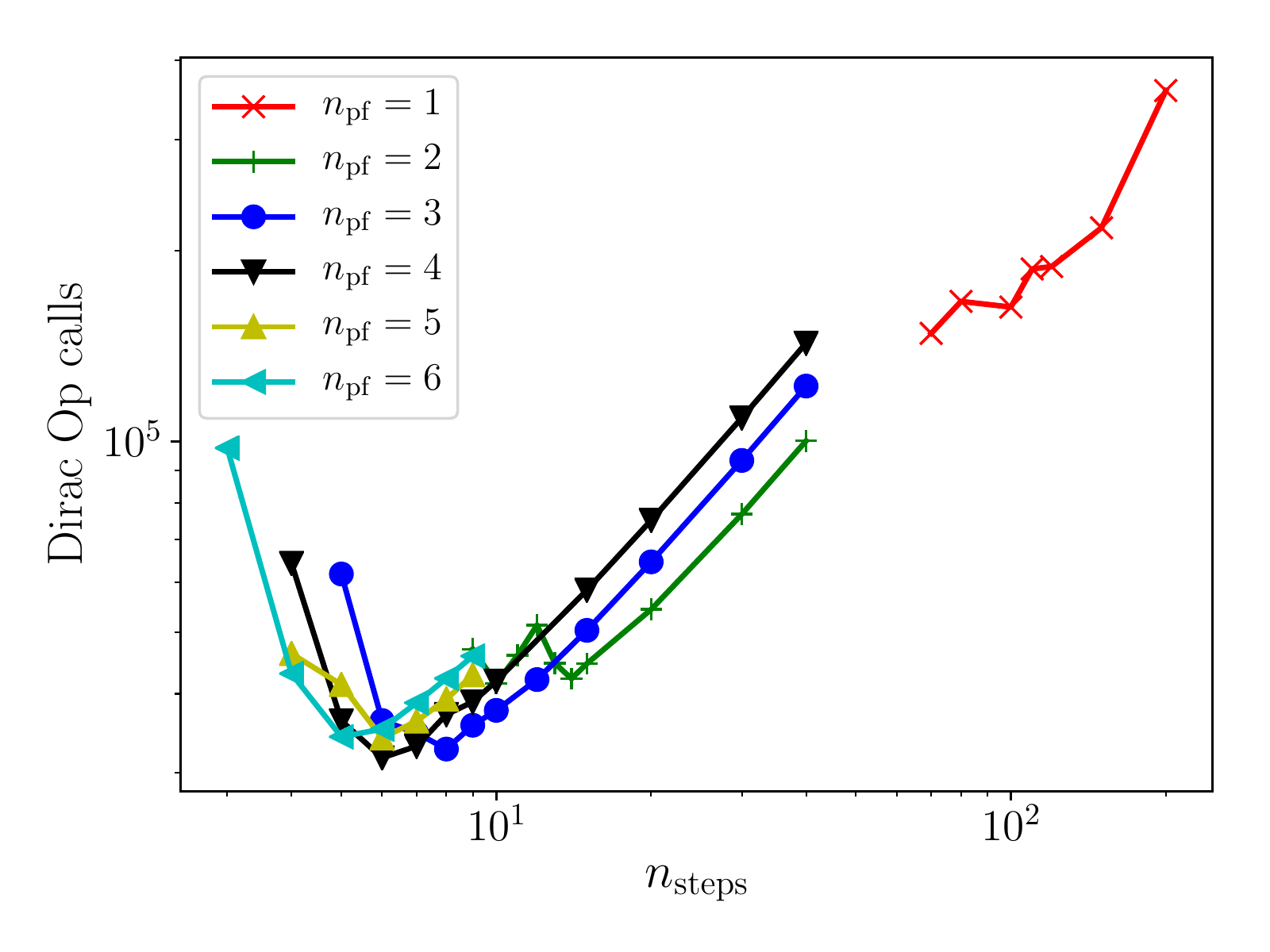}\\\includegraphics[width=25em]{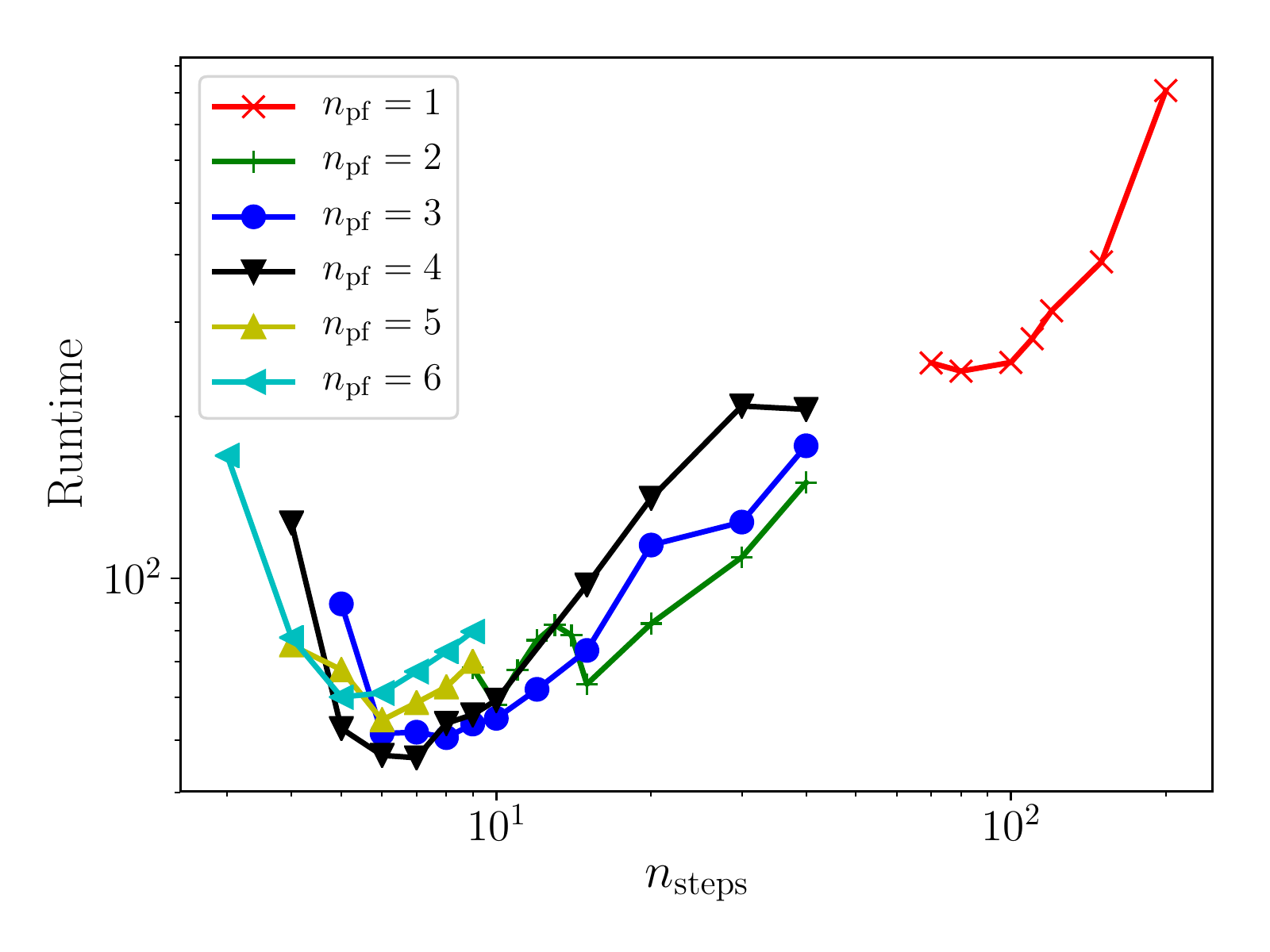}
\caption{The cost of generating an accepted $\tau=1$ trajectory using the block multishift (SBCGrQ) inverter for different $\npf$ versus the number of integrator steps $n_{\mathrm{steps}} = \tau/\delta\tau$. \emph{Top}: cost in Dirac operator calls, \emph{bottom}: computer runtime cost.}
\label{fig:runs_cost}
\end{center}
\end{figure}

So far we have compared solvers for different $\npf$ while keeping the residual of the lowest shift the same, but from Eq.~(\ref{eq:convergence_bound_shifts}) we can also expect the residuals of the shifted solutions to depend on $\npf$. Fig.~\ref{fig:residuals} shows the residual of shifted solutions using the SBCGrQ solver (for $\npf=1$ this reduces to the SCG solver), for a wide range of shifts $\sigma$.
In the top panel, the number of solver iterations $k$ is kept constant, and we see the residuals for small shifts decrease dramatically as $\npf$ is increased, which is consistent with the expectation from Eq.~(\ref{eq:convergence_bound}). In the bottom panel, the number of solver iterations is adjusted such that the unshifted relative residual is $\norm{r}/\norm{r_0}\simeq10^{-7}$. Here we see a relative increase in the shifted residuals for intermediate shifts, as predicted by Eq.~(\ref{eq:convergence_bound_shifts}), since fewer iterations are required as $\npf$ is increased. For large values of $\npf$ this might mean that a tighter residual for the force term inversions will be required to maintain the accuracy of the force term, but we saw no such issues in our runs for $\npf\leq 6$ where we use the same stopping criterion for all $\npf$.

\subsection{Combined Results}

Combining our results from the previous two sections we can measure the cost of generating an accepted RHMC trajectory in two ways. One is in terms of Dirac operator calls per trajectory divided by the acceptance rate, which is implementation--independent but does not take into account the acceleration of the Dirac operator or the overhead of the multishift block solver. The second measure of the cost is simply the CPU time required by our reference implementation (running on a single thread of a CPU) to generate a trajectory, divided by the acceptance rate. This takes all the costs into account, but the results are now heavily implementation--dependent, and as our implementation is not parallelised and prioritises flexibility over performance the results may be significantly different on a fully optimised production lattice QCD code. Moreover, GPU--based hardware with a higher ratio of compute performance to memory bandwidth should benefit more from the increased arithmetic intensity of the block Dirac operator.

\begin{figure}
\begin{center}
  \includegraphics[width=25em]{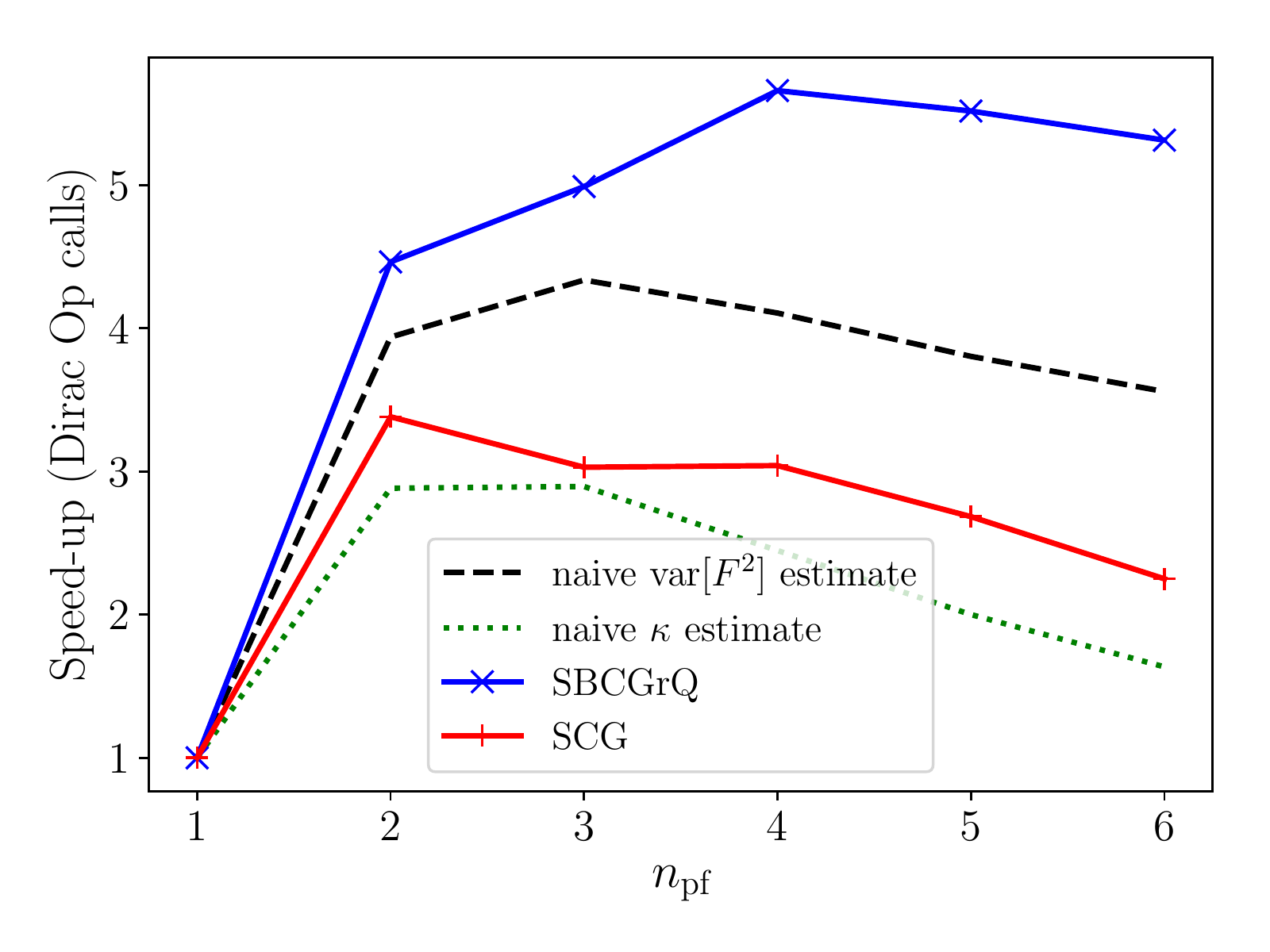}\\\includegraphics[width=25em]{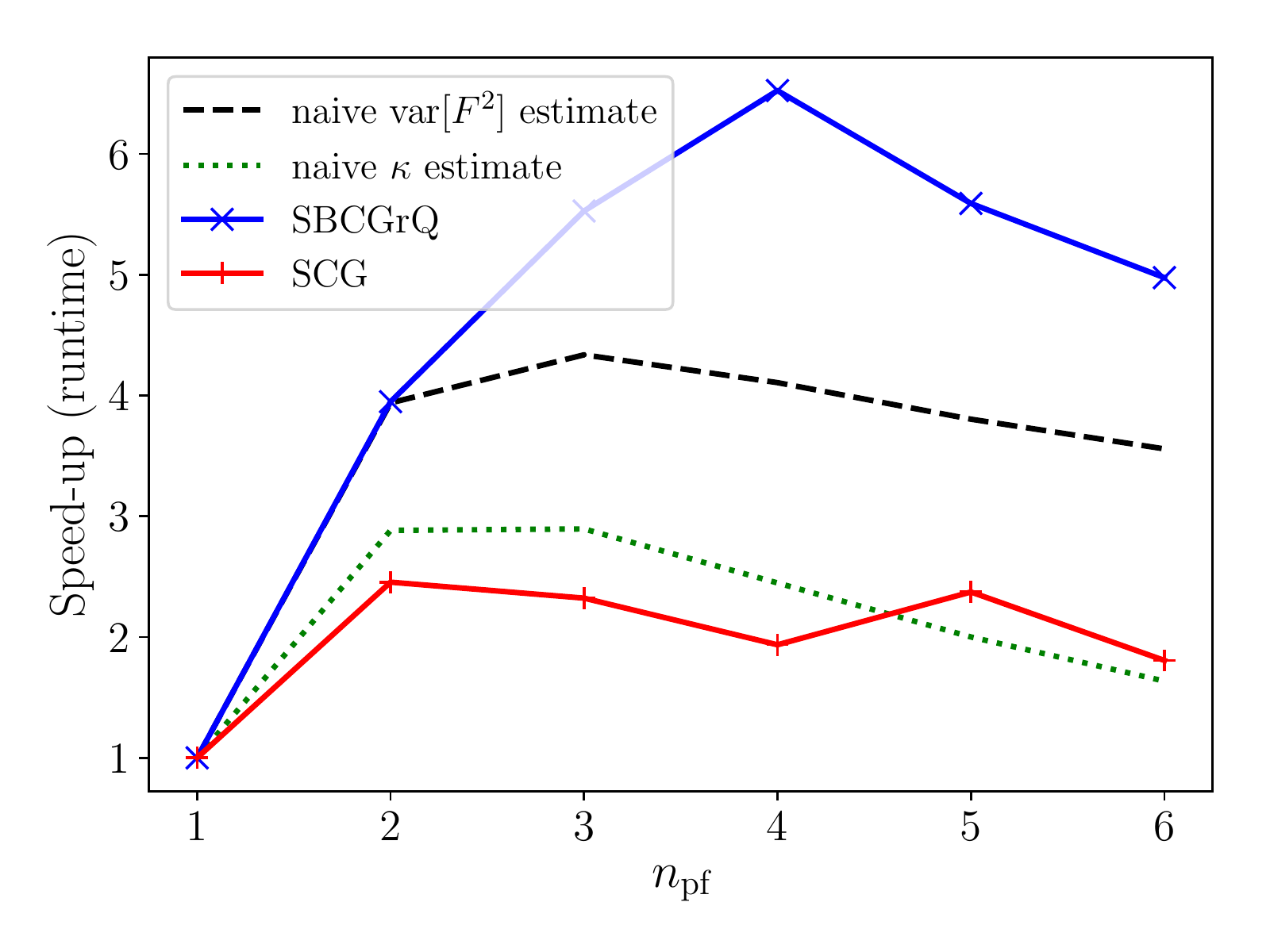}
\caption{Trajectory speed-up versus $\npf$ at $\simeq90\%$ acceptance, normalised to 1 for $\npf=1$, using either multishift (SCG) or block multishift (SBCGrQ) solvers. \emph{Top}: speed-up in Dirac operator calls, \emph{bottom}: speed-up in computer runtime. The black dashed line is the simple prediction using the force norm variance of Eq.~(\ref{eq:cost}), and the green dotted line is the simple prediction from the condition number of the Dirac operator using Eq.~(\ref{eq:cost_clark}).  The optimal $\npf$ and the overall gain are both significantly increased by the use of block methods.}
\label{fig:cost}
\end{center}
\end{figure}

Both measures of the cost are shown in Fig.~\ref{fig:runs_cost} as a function of the integrator step size for $\npf=1$ to $6$, using the SBCGrQ inverter and block Dirac operator. For both cost measures there is a clear benefit from increasing $\npf$ to $3$ or $4$. The optimal integrator step size for each $\npf$ in this plot corresponds to a $\simeq90\%$ acceptance rate. Taking these optimal integrator step sizes we can compare the overall improvement the block method offers compared to the previous non--block results of Sec.~\ref{sec:results:npf}, which is shown in Fig.~\ref{fig:cost}. We see a $\sim 6\times$ speed-up using $\npf=4$ compared to HMC, while the non--block multishift CG solver gave a $\sim 3\times$ speed-up using $\npf=2$.

\section{Conclusions}
\label{sec:conclusions}
Let us summarize our study. We find that using multiple, $\npf > 1$, 
pseudofermions in RHMC simulations of lattice QCD offers three {\em cumulative} advantages:
\begin{enumerate}
\item The magnitude of the fermionic force is reduced, which allows an increase
of the integrator step size. Fewer steps are required per trajectory.
\item The computation of the pseudofermionic force at each step now involves
solving $\npf$ linear systems with different right hand sides, all with 
the same Dirac matrix. Such systems are advantageously solved by {\em block}
Krylov solvers, which converge with {\em fewer} Dirac matrix-vector operations, 
because the dimension of the search Krylov space increases by $\npf$ at each iteration.
\item The computing time for a Dirac matrix-vector operation decreases, because
the gauge field entering the Dirac matrix needs only to be loaded once for 
$\npf$ vectors to be multiplied, and cache locality is improved.
\end{enumerate}
In addition, one may speculate that a smaller fermionic force, as obtained
by multiple pseudofermions, indicates a smoother energy landscape, which 
might be explored faster by RHMC dynamics. We looked for a possible reduction
of autocorrelation time under an increase of $\npf$, but found no 
clear indication of such (see Table~\ref{tab:long_runs}).

The solver that we use, described in Algorithm~\ref{alg:SBCGrQ}, is a multishift block version
of the conjugate gradient, constructed in Ref.~\cite{futamura}. The problem of numerical instability seen in previous
block solvers is handled by re-orthogonalization of the search matrix, as 
recommended in Ref.~\cite{Dubrulle2001} and recently used in Refs.~\cite{Nakamura:2011my,Clark:2017ekr}.

Our simulations, albeit on a small lattice, show that 3 or 4 pseudofermions
allow for a gain $\mathcal{O}(6)$ in CPU time. Let us discuss what to expect in
a more realistic setup.

An improved, less local Dirac operator of staggered type would probably lead to 
further CPU gains because the assembly of the Dirac matrix elements from memory
could be amortized even better. Similarly, a GPU-type architecture would
benefit more, since its memory bandwidth is typically more limited compared
to its FLOP performance.
Ref.~\cite{Clark:2017ekr} has shown significant gains from a block solver on
a GPU machine. The multishift version thereof should yield similar
benefits.

The reduction in solver iterations is strongly dependent on the ratio of the $\npf$-th eigenvalue of the Dirac operator to the smallest one - the larger this ratio the greater the reduction in the number of iterations, as predicted from the convergence bound of Eq.~(\ref{eq:convergence_bound}) and also as seen empirically in Fig.~\ref{fig:convergence}. This observation can guide our expectations for how the gain from the block solver should depend on the mass, volume and lattice spacing. In general, reducing the mass, going to coarser lattice spacing or reducing the physical volume should all increase the gain of the block solver. Conversely increasing the mass, going to finer lattice spacing or increasing the physical volumes would presumably reduce the benefits of the block solver, so one scenario where this method may be particularly advantageous would be simulations done in the $\epsilon$-regime.

The benefit from using multiple pseudofermions in the molecular dynamics also grows as the mass is reduced, moreover the reduced variance of the force term would allow the use of higher order (but less stable) integrators whose costs grow more slowly with the volume~\cite{Clark:2006fx}.

A more quantitative statement about the scaling of the method with these parameters and how it compares to other recent algorithmic improvements such as multigrid~\cite{Frommer:2013fsa,Brower:2018ymy} and deflation~\cite{Luscher:2007se} would be highly desirable, but would require large scale simulations that are beyond the scope of this work.

Finally, we emphasize that our approach is algorithmically simple;
more realistic tests involve rather small amounts of programming, and a single
parameter to optimize: the number $\npf$ of pseudofermions.

\begin{acknowledgments}
This work is supported by the Swiss National Science Foundation under the grant 200020-162515. Numerical simulations were performed on the Euler cluster at ETH Z\"urich. The authors thank the CERN Theoretical Physics Department for its hospitality. 
\end{acknowledgments}
  
\bibliographystyle{utphys}
\bibliography{rhmc_blockCG.bib}

\end{document}